\documentclass{emulateapj}
\usepackage{lscape}
\newcommand{\balq}{BAL~quasar}

\newcommand{\balqs}{BAL~quasars}

\newcommand{\chandra}{{\emph{Chandra}}}

\newcommand{\xmm}{\emph{XMM-Newton}}

\newcommand{\iras}{{\emph{IRAS}}}
\newcommand{\spitzer}{{\emph{Spitzer}}}

\newcommand{\kms}{\mbox{\,km\,s$^{-1}$}}
\newcommand{\cmsq}{\mbox{\,cm$^{-2}$}}

\newcommand{\fnu}{\mbox{\,erg\,cm$^{-2}$\,s$^{-1}$\,Hz$^{-1}$}}
\newcommand{\lumin}{\mbox{\,erg~s$^{-1}$}}
\newcommand{\persec}{\mbox{\,s$^{-1}$}}
\newcommand{\nh}{\mbox{${N}_{\rm H}$}}

\newcommand{\CIV}{\ion{C}{4}}

\newcommand{\OVI}{\ion{O}{6}}

\newcommand{\altcite}{\citealt}

\newcommand{\total}{38}

\newcommand{\microns}{\micron}
\newcommand{\leight}{$l_{\rm 8\mu{m}}$}
\newcommand{\lfive}{$l_{\rm 5000\AA}$}
\newcommand{\ltwo}{$l_{\rm 2500\AA}$}
\newcommand{\alphao}{$\alpha_{\rm o}$}
\newcommand{\lsun}{$L_{\odot}$}
\begin{document}
 
 
\shortauthors{Gallagher et al.}
\shorttitle{Radio through X-ray SEDs of BAL Quasars}

\title{Radio Through X-ray Spectral Energy Distributions of \total\ Broad
Absorption Line Quasars}

\author{S.\ C. Gallagher,\altaffilmark{1}
D.\ C. Hines,\altaffilmark{2}
Myra Blaylock,\altaffilmark{3}
R.\ S. Priddey,\altaffilmark{4}
W.\ N. Brandt,\altaffilmark{5}
\& E.\ E. Egami\altaffilmark{3}
}

\altaffiltext{1}{Department of Physics \& Astronomy, University of
  California -- Los Angeles, 430 Portola Plaza, Box 951547, Los
  Angeles CA, 90095--1547, USA; {\em sgall@astro.ucla.edu}}
\altaffiltext{2}{Space Sciences Institute, 4750 Walnut Street, Suite 205
Boulder, CO 80301, USA}
\altaffiltext{3}{Steward Observatory, University of Arizona, 933 North
Cherry Avenue, Tucson, AZ 85721, USA}
\altaffiltext{4}{Centre for Astrophysics Research, University of
        Hertfordshire, College Lane, Hatfield, Hertfordshire AL10 9AB,
        UK}
\altaffiltext{5}{Department of Astronomy and Astrophysics, The
        Pennsylvania State University, University Park, PA 16802, USA}
       
\begin{abstract}
We have compiled the largest sample of multiwavelength spectral energy
distributions (SEDs) of Broad Absorption Line (BAL) quasars to date,
from the radio to the X-ray.  We present new \spitzer\ MIPS (24, 70,
and 160\micron) observations of \total\ \balqs\ in addition to data
from the literature and public archives. In general, the mid-infrared
properties of \balqs\ are consistent with those of non-\balqs\ of comparable
luminosity.  In particular, the optical-to-mid-infrared luminosity
ratios of the two populations are indistinguishable.  We also measure
or place upper limits on the contribution of star formation to the
far-infrared power.  Of 22 (57\%) upper limits, seven quasars have
sufficiently sensitive constraints to conclude that star formation
likely contributes little ($<20\%$) to their far-infrared power.  The
17 \balqs\ (45\%) with detected excess far-infrared emission likely
host hyperluminous starbursts with $L_{\rm
fir,SF}=10^{13-14}$\lsun. Mid-infrared through X-ray composite \balq\
SEDs are presented, incorporating all of the available photometry.
Overall, we find no compelling evidence for inherent differences
between the SEDs of BAL vs. non-\balqs\ of comparable luminosity.  Therefore
a ``cocoon'' picture of a typical \balq\ outflow whereby the wind covers
a large fraction of the sky is not supported by the mid-infrared SED
comparison with normal quasars, and the disk-wind paradigm
with a typical radio-quiet quasar hosting a BAL region remains viable.
\end{abstract}


\section{Introduction}
\label{sec:intro}

Recent theoretical investigations into quasar feedback as a potentially
important cosmological force in galaxy formation
\citep[e.g.,][]{ScOh2004,GranatoEtal2004,SpDiHe2005,hopkins+05} have
led to a expansion of interest in quasar outflows.  Energetic outflows
are most obviously manifested in Broad Absorption Line (BAL) quasars.
This population, approximately 20\% (once selection effects are taken
into account) of optically selected type~1 quasar samples
\citep[e.g.,][]{HewFol2003,reich+03a}, is notable for broad,
blueshifted absorption evident in common ultraviolet resonance
transitions such as
\CIV, Ly$\alpha$, and \OVI.  These P~Cygni-type features arise because
the observer is looking through an outflowing wind.  Both the high
velocities (up to 0.03 to 0.1$c$) and the relatively high ionization
state of the gas indicate that the wind originates within the inner
parsec of the central engine, and is likely more or less co-spatial
with the broad-line region gas \citep{MuChGrVo1995,PrStKa2000}.

While the line of sight to an individual \balq\ certainly cuts through
an outflowing wind, the average covering fraction of the outflow
remains difficult to constrain.  For example, it may be that \balqs\
have a wind that covers a large fraction of the sky, and therefore
only about 20\% of quasars have BAL outflows.  Alternately, most
quasars may host BAL outflows, with the 20\% (corrected) population
fraction resulting from the average BAL wind covering fraction.  The latter
interpretation is supported by evidence from spectropolarimetry that
there are absorption-line-free lines of sight in \balqs\
\citep[e.g.,][]{OgCoMiTr1999}, and the wind covering fraction is
inferred to be low from constraints on ultraviolet emission-line
scattering \citep{HaKoMo1993}. \citet{WeMoFoHe1991} found an overall
similarity of BAL and non-BAL quasar ultraviolet-optical continuum and
emission-line properties.  More subtly, the significantly larger
spectral comparison of \citet{reich+03b} revealed that \balqs\ are
preferentially drawn from the non-\balq\ population with intrinsically
bluer continua, though they concluded that the parent populations of
optically selected BAL and non-\balqs\ appear to be the same.  At
submm wavelengths, where the thermal dust emission is generated on
much larger scales than the accretion disk, the populations of BAL and
non-\balqs\ have indistinguishable flux distributions, though the
majority of both populations have only upper limits
\citep{lewis+03,willott+03,priddey+06}.\footnote{\citet{priddey+06}
also found the intriguing result that of their 16 \balqs, the
submm-detected objects tended to have the largest \ion{C}{4}
absorption EW.} These empirical results are consistent with the
disk-wind paradigm whereby a radiatively driven, equatorial wind from
the accretion disk gives rise to both the broad emission and broad
absorption-line features \citep{MuChGrVo1995}.  However, a
near-equatorial line of sight has been called into question by radio
observations where typical orientation indicators (e.g., steep
vs. flat radio spectra, and core vs. lobe-dominated flux ratios)
suggest a range of inclination angles for radio-loud \balqs\
\citep{BeckerEtal2000,zhou+06,brotherton+06}. Furthermore, the
identification of a few \balqs\ with ultra-luminous infrared galaxies
\citep[ULIRGS; e.g.,][]{CaSt2001} has led to the interpretation that
\balqs\ may be ``cocooned'' during an obscured evolutionary stage in the transition from
galaxy merger to starbursting ULIRG to shrouded \balq\ to naked quasar
\citep[e.g.,][]{SaEtal1988,gregg+02}.

The apparently contradictory results of these studies have led us to
undertake a systematic survey of a well-defined sample of \balqs\
drawn from the Large Bright Quasar Survey
\citep[LBQS;][]{lbqs_ref,HewFol2003}.  The goal is to obtain
multiwavelength coverage of a sizable number of
\balqs\ for comparison of the spectral energy distributions (SEDs) of
BAL and non-\balqs\ in an effort to investigate the underlying
relationship between the two populations.  To date, there are two
notable differences between BAL and non-\balq\ SEDs: (1) \balqs\
typically show more evidence for dust reddening and extinction in their
ultraviolet-optical spectra \citep{SpFo1992,ric+02,reich+03b,trump+06}, and
(2) \balqs\ are remarkably weak X-ray emitters
\citep[e.g.,][]{GrMa1996,gall+06}.  However, both of these properties are
consistent with orientation-dependent obscuration effects. 

In this present study, we focus on incorporating near and mid-infrared
photometry into the picture, as longer wavelength data are expected to
be less sensitive to absorption and less influenced by orientation
effects.  In particular, we will test the conclusion of
\citet{KrVo1998} that limb-darkening of the accretion disk emission
should lead \balqs\ to have relatively weak ultraviolet-optical
continua compared to non-\balqs\ if the BAL outflow is strongly
equatorial.  \citet{Goodrich1997} reached a similar conclusion, though
the proposed mechanism was orientation-dependent attenuation (perhaps
via electron scattering) based on polarization studies.  The
ultraviolet-optical emission in quasars is generated by the accretion
flow near the black hole, while infrared continuum light longward of
$\sim1$\,\micron\ is generally attributed to a parsec-scale, cold, and
dusty region that is heated by and reprocesses the higher energy,
direct continuum.  Therefore, the mid-infrared power in \balqs\ should
provide an independent measure of the intrinsic accretion luminosity
to test these hypotheses.

Throughout we assume $\Lambda$-CDM cosmology with $\Omega_{\rm
M}=0.3$, $\Omega_{\Lambda}=0.7$, and $H_0=70$\kms\,Mpc$^{-1}$
\citep{sperg03,sperg06}.

\section{Sample Selection}

This sample was drawn from the Large Bright Quasar Survey
\citep{lbqs_ref} sample of \balqs\ identified by
\citet{HewFol2003}.  Each object has a BAL probability of 1.0 and
$z>1.4$, the redshift at which the definitive C IV BAL is shifted into
the wavelength regime accessible with ground-based spectroscopy.  
Of the 44 total \balqs\ meeting these criteria identified by
\citet{HewFol2003}, 38 were chosen for \spitzer\ 24, 70, and 160\micron\
MIPS observations; we present this MIPS sample in this paper.  The
MIPS targets were chosen to maximize the available multiwavelength
coverage.  In addition to the mid and far-infrared coverage, available
data archives and literature were searched to augment the
multiwavelength photometry as much as possible. The targets are
optically bright with $B_J$\footnote{This photographic blue magnitude
is related to the more standard Johnson blue magnitude: $B_J = B -
0.28(B - V)$ \citep{lbqs_ref}.} magnitudes of 16.68--18.84, and they
are drawn from a homogeneous, magnitude-limited quasar survey that has
effective, well-defined, and objectively applied selection criteria
\citep[e.g.,][]{HeFoCh2001}; six known \balqs\ with low-ionization
(\ion{Mg}{2} and/or \ion{Al}{3}) BALs (hereafter, LoBALs) are
included. The overall sample is luminous, with $M_B$ of --26.2 to
--28.3 (using the K-corrections for non-BAL quasars of
\altcite{CrVi1990}).

Our observed MIPS sample is listed in Table~\ref{tab:lums} including
the redshift and BAL type.  The classifications are HiBAL (no
\ion{Al}{3} or \ion{Mg}{2} BAL detected), LoBAL (\ion{Mg}{2} BAL
detected), and Unknown (the electronic spectra do not cover \ion{Mg}{2}).

\section{Multiwavelength Photometry}

The foundation of this multiwavelength SED project is complete
mid and near-infrared photometry from \spitzer\ MIPS and the Two
Micron All Sky Survey \citep[2MASS;][]{2mass_ref} of \total\ LBQS
\balqs.  These data have been supplemented by near-complete (35 of
\total) 0.5--8.0~keV X-ray data, multiband optical photometry for all,
submm coverage for 16 objects, and Very Large Array (VLA) radio observations of 35 out of
\total. In this section, the provenance of the SED data is briefly
described.

\subsection{\spitzer\ MIPS Photometry} 
\label{sec:mips}

The bulk of the MIPS data is the 36 \balq\ observations from the
Cycle~1 Director's Discretionary Time for Gallagher's \spitzer\
Fellowship program.  Two additional objects, 0059$-$2735 and
1331$-$0108, are added from Program \#82 (PI Rieke) of Cycle 1
Guaranteed Time.

The MIPS observation modes were one 3 s cycle at 24\micron\ (48 s
integration), one 3 s or 10 s cycle at 70\micron\ (38 or 128 s
integration), and four 10 s cycles at 160\micron\ (84 s integration).
The longer 70\micron\ integration times were chosen for fainter
sources, and the 160\micron\ integration times were chosen to reach
the confusion limit.  All \balqs\ were detected with high S/N at
24\micron.  We consider \balqs\ with S/N$>3$ to be detected at 70 and
160\micron.  The upper limits are calculated as the flux in the source
aperture + 3$\sigma$ or 3$\sigma$, whichever is greater (following the
convention of the MIPS instrument team), where $\sigma$ is the
standard deviation in the mean background.  By this criterion, 13
\balqs\ are detected at 70\micron, and three are detected at
160\micron.  (The two Guaranteed Time targets do not have 160\micron\
coverage.)  The \spitzer\ photometry is presented in
Table~\ref{tab:phot1}.

\subsection{Near-infrared}
\label{sec:2mass}

Thirty-two out of \total\ \balqs\ are included in the 2MASS Point
Source Catalog \citep[PSC;][]{2mass_ref}; of these, all had $JHK$
detections except for 00211$-$0213, 1133$+$0214, 1314$+$0116, and
2116$-$4439, which only have upper limits in $K$.  To obtain
photometry (either upper limits or detections) for the remaining six
(0051$-$0019, 1208$+$1535, 1240$+$1607, 1243$+$0121, 1443$+$0141, and
2140$-$4552), we downloaded the best Atlas $JHK$ images (recommended
for photometry) from the 2MASS
database.\footnote{http://irsa.ipac.caltech.edu/applications/2MASS/IM/}
These images have all been flux-calibrated, combined, and resampled to
a 1\arcsec/pixel scale; the photometric zeropoints are included in the
fits image headers.  We measured the signal (in data numbers) within a
2\arcsec-radius circular source cell centered on the LBQS optical
position; the background was determined from an encircling annulus
with inner and outer radii of 8\arcsec\ and 10\arcsec,
respectively. The noise per pixel, $\sigma$, is the standard deviation
in the mean background. If the S/N within a 2\arcsec\ aperture was
more than 2.5, we consider the source to be detected.  Otherwise, the
upper limit is set to the 3$\sigma$ background level within the source
cell area.

While a 2\arcsec\ source cell maximized the S/N, a significant
fraction of the flux in the point spread function is distributed
outside of this aperture.  To determine the aperture correction (which
varies from field to field and band to band), we performed aperture
photometry on the 2MASS PSC catalog sources with `AAA' photometric
quality flags and no galaxy contamination within 300\arcsec\ of the
target that coincided with the Atlas image area.  The aperture
correction for each target in each band was then the mean difference
between the catalog magnitudes and their corresponding
2\arcsec-aperture magnitudes.  The number of point sources used in the
aperture correction determination was between four and 16; aperture
correction values ranged from $-0.42$ to $-0.57$ magnitudes.  The aperture
correction was added to both the detections and upper limits to give
the fluxes (converted to mJy from magnitudes) listed in
Table~\ref{tab:phot2}.  The standard deviation in the mean aperture
correction has been folded into the photometric error.

Following the procedure outlined above, we obtain 37/35/32 detections
in the $J$/$H$/$K$ bands, respectively, for the \total\ objects in the
sample.

\subsection{X-ray}
\label{sec:xray}

Thirty-four of the \total\ objects in the sample were observed
with \chandra, and a detailed analysis of these data and comparison
with ultraviolet spectroscopic properties is presented in
\citet{gall+06}.  In addition, 2212$-$1759 was observed for 172~ks
with \xmm\ \citep{clavel+06}.  Using the 0.5--2.0~keV EPIC-pn
count-rate upper limit ($<9.25\times10^{-5}$ count\persec), and
assuming a power-law spectral index of $\alpha_{\rm X}=0.0$
(appropriate for absorbed sources), the modeling tool
WebPIMMS\footnote{http://heasarc.nasa.gov/Tools/w3pimms.html} gives
$f_{\rm 2keV}<4.12\times10^{-34}$\fnu.  

In Table~\ref{tab:phot2}, the observed-frame 1~keV photometry is
listed in units of $10^{-6}$~mJy for the 35 sources with X-ray data.
The asymmetric flux errors result from Poisson X-ray counting
statistics.

\subsection{Optical}
\label{sec:sdss}

We have cross-correlated the LBQS \spitzer\ \balq\ sample with the
SDSS Data Release 5 archive.  Of the \total\ MIPS-observed \balqs, 22
have available SDSS photometry.  We list the dereddened (for Galactic
extinction) $ugriz$ PSF magnitudes for these objects in
Table~\ref{tab:phot2} converted to units of mJy using the SDSS
zeropoint of 3631 Jy.

To obtain constraints on the optical colors of the remaining quasars,
we convolved the electronic spectra (from \altcite{KoVoMoWe1993}
[preferred] or \altcite{lbqs_ref}) with the SDSS $gri$ filter
functions for those quasars without SDSS coverage after first
normalizing the available spectra using the LBQS $B_J$ photometry and
transmission curves \citep[Appendix A;][]{bj_ref}.  These synthetic
broad-band fluxes are presented in Table~\ref{tab:phot2} (in units of
mJy), errors are propagated from the 0.15 mag $B_J$ uncertainty
\citep{lbqs_ref}.  The quoted uncertainty does not include additional
errors from the convolution. The number of bandpasses is limited to
the wavelength coverage of the electronic spectra.

To check on the absolute photometric accuracy of the synthetic fluxes,
synthetic SDSS photometry was calculated for those quasars with actual
SDSS data using non-SDSS electronic spectra. In sum, the average
offsets between the synthetic and SDSS $gri$ photometry (defined as
(SDSS -- synthetic)/SDSS [mJy]) are within the 15\% photometric
uncertainty of the LBQS. The mean offsets and standard deviations of
the mean offsets for $g$, $r$, and $i$ are $0\pm27\%$, $-5\pm37\%$,
and $-12\pm33\%$, respectively.  The large standard deviations likely
indicate that the synthetic photometric errors are underestimated,
however, there is no evidence for absolute, systematic offsets between
the synthetic and SDSS photometry.

\subsection{Submillimeter}
\label{sec:scuba}

Sixteen of the \total\ \balqs\ have been observed at 450 and
850\microns\ with the Submillimetre Common-User Bolometer Array
(SCUBA) instrument at the James Clerk Maxwell Telescope. The SCUBA
data reduction and analysis is described in detail in
\citet{priddey+06}.  In this paper, we use $S/N\ge2.5$ as the
detection criterion; at 850\micron\ (450\micron), this yields five
(two) detections with a 1$\sigma$ limit of $\sim1.5$~mJy.  A more
conservative 3$\sigma$ threshold reduces the number of 850\micron\
detections to three.  For non-detections, the upper limits are given
as the flux in the source aperture + $2\sigma$ or $2\sigma$,
whichever is greater.  The available photometry is included in
Table~\ref{tab:phot1}.

\subsection{Radio}
\label{sec:radio}

Radio coverage is publicly available for 35 of \total\ \balqs.  These
data are compiled from \citet{stocke+92} (5~GHz), the Faint Images of
the Radio Sky at Twenty cm \citep[FIRST;][]{first_ref} (1.4~GHz), and
the National Radio Astronomical Observatory VLA Sky Survey
\citep[NVSS;][]{nvss_ref} (1.4~GHz).  Data are listed in
Table~\ref{tab:phot1}.  Upper limits for the FIRST survey are as given
for the field in the catalog search; NVSS upper limits are 5~mJy
\citep{nvss_ref}.

Only one of the quasars in our sample, 2211$-$1915, is known to be
formally radio-loud as defined by a radio luminosity of
$>10^{32}$\lumin~Hz$^{-1}$. The three quasars without radio constraints are too
far south for the VLA. Based on the radio-loud fraction of the LBQS
determined from a cross-correlation with the FIRST survey \citep{HeFoCh2001},
we expect at most 12\% ($<1$) of these to be radio-loud, although a
more likely value is 6\% based on the lower fraction of radio-detected
BAL versus non-\balqs\ \citep{HewFol2003}.

\section{Results and Discussion}
\label{sec:seds}

\begin{figure*}
\figurenum{1a}
\plotone{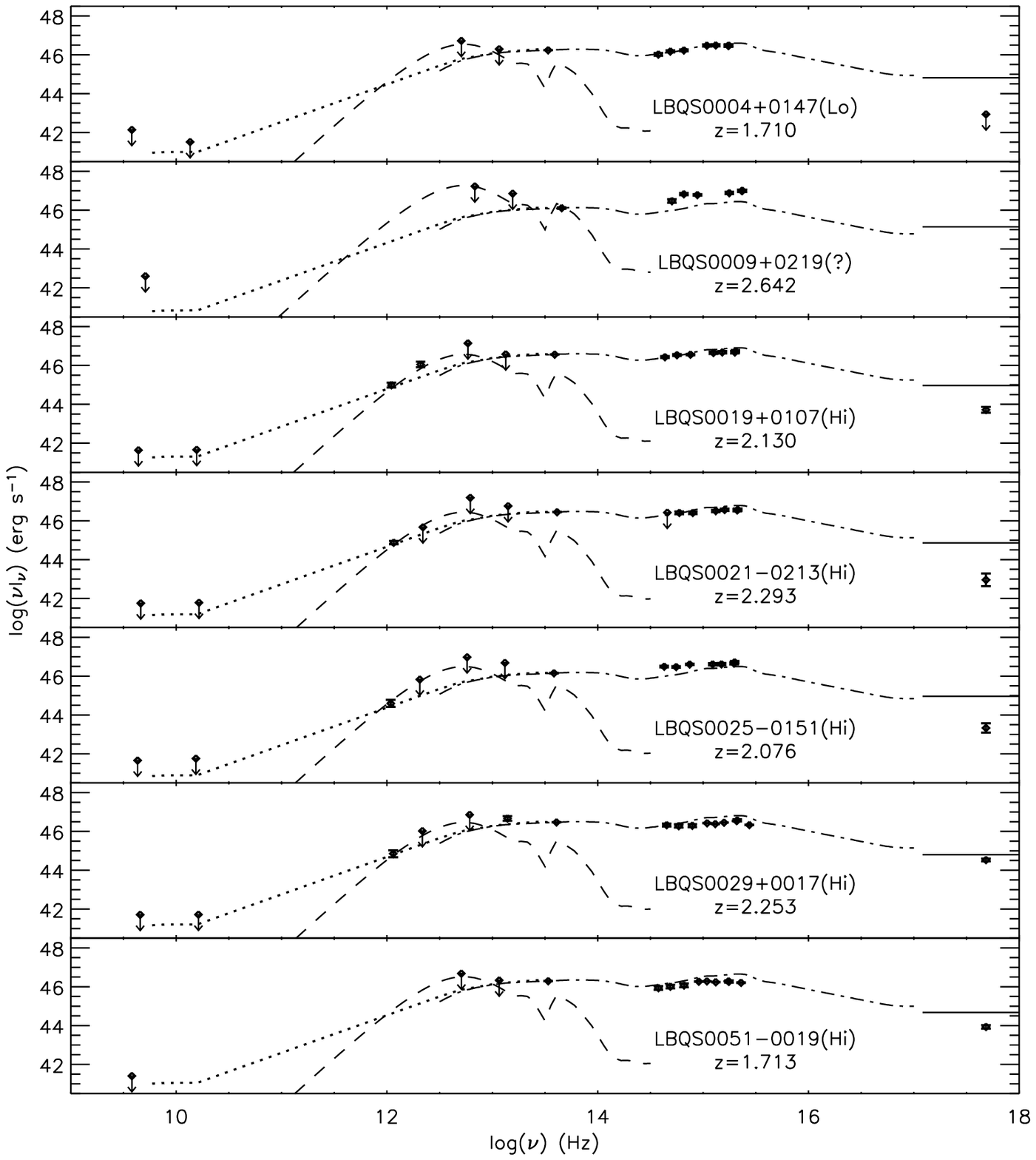}
\caption{ SEDs for the first seven \balqs\ listed in
Table~\ref{tab:lums}; data are listed in Tables~\ref{tab:phot1} and
\ref{tab:phot2}.  The composite SEDs (dotted from
\altcite{ElvisEtal1994}, dot-dashed from \altcite{ric+06}) have been
normalized to the 24\micron\ data point.  The predicted X-ray SED
(solid line) is normalized from the 2500\AA\ monochromatic luminosity
(see \S\ref{sec:seds}). The dashed curve is a model starburst SED
normalized to the far-infrared photometry in excess of the composite
quasar SED (see \S\ref{sec:fir}). Objects are labeled with name, BAL
type, and redshift.}
\label{fig:sed1}
\end{figure*}

\begin{figure*}
\figurenum{1b}
\plotone{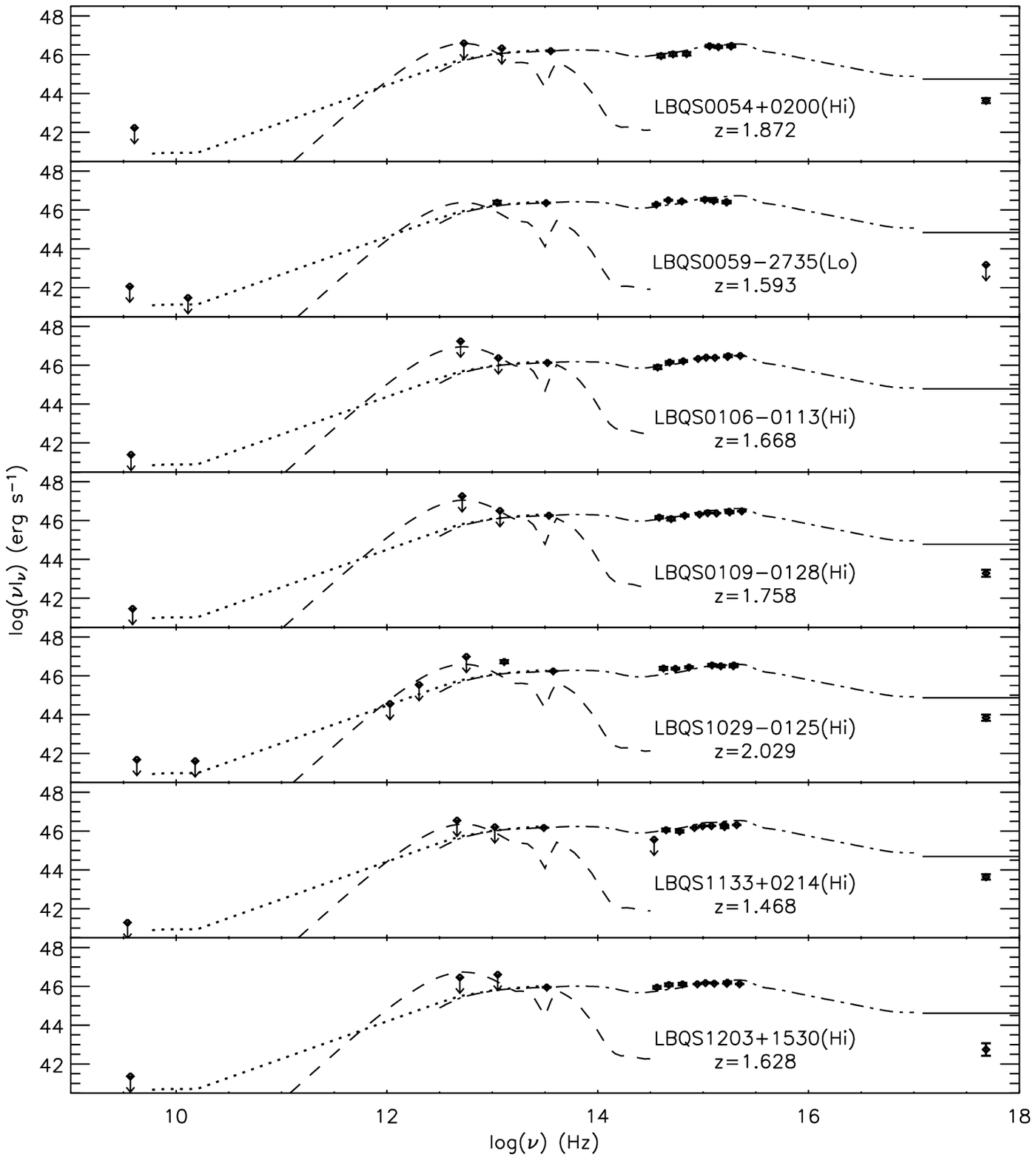}
\caption{
See Figure~\ref{fig:sed1} caption.}
\label{fig:sed2}
\end{figure*}

\begin{figure*}
\figurenum{1c}
\plotone{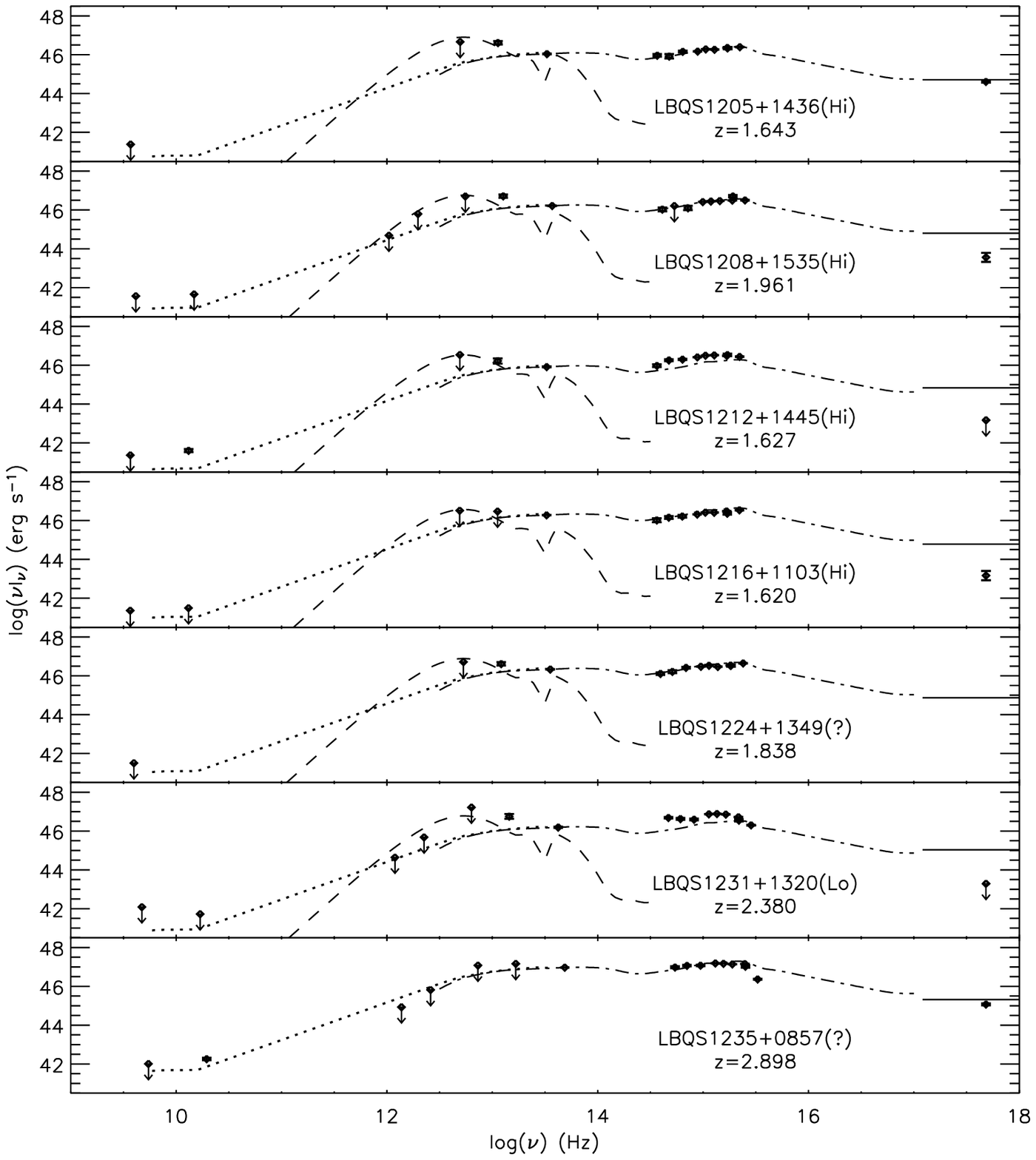}
\caption{
See Figure~\ref{fig:sed1} caption.}
\label{fig:sed3}
\end{figure*}

\begin{figure*}
\figurenum{1d}
\plotone{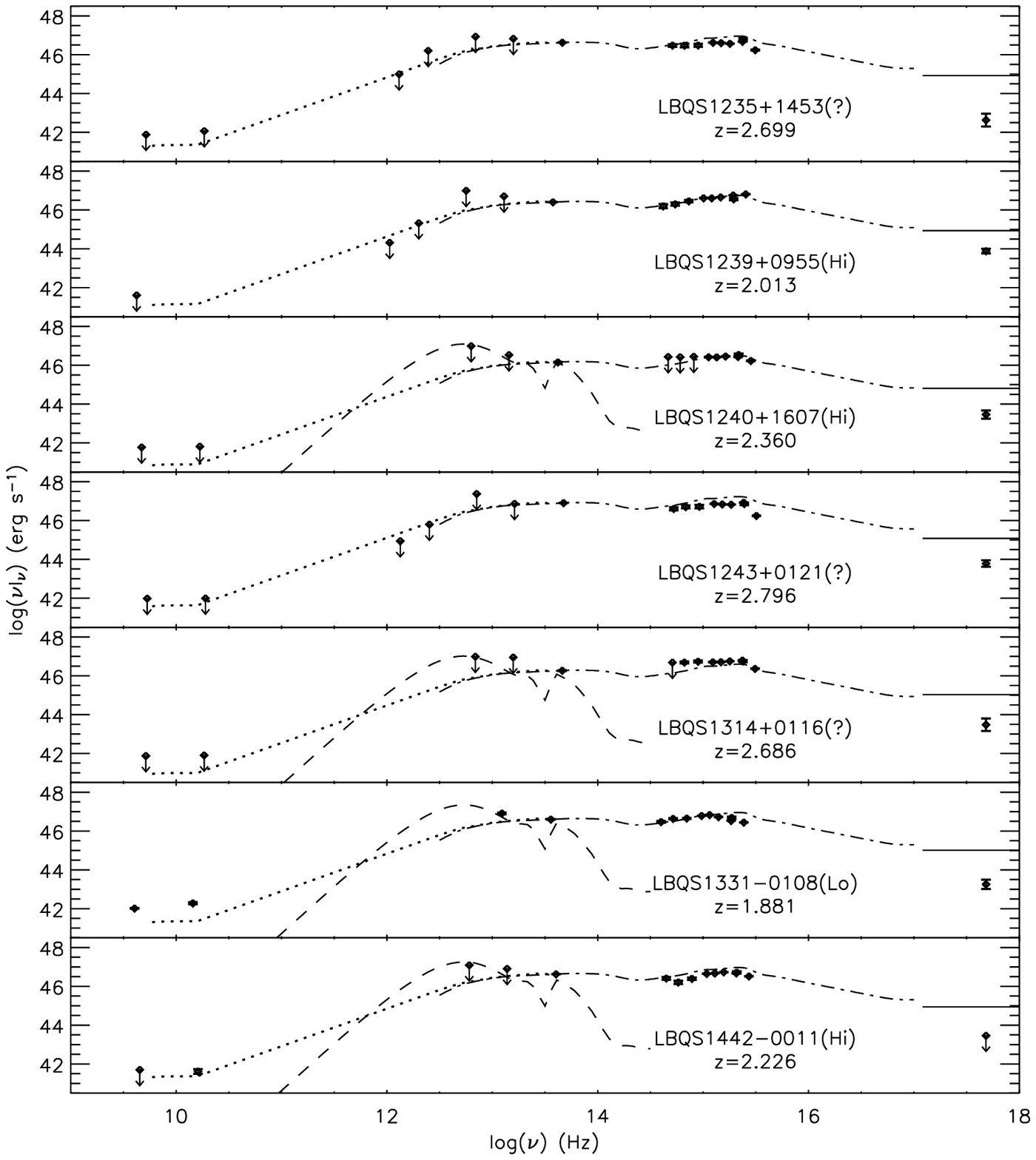}
\caption{
See Figure~\ref{fig:sed1} caption.}
\label{fig:sed4}
\end{figure*}

\begin{figure*}
\figurenum{1e}
\plotone{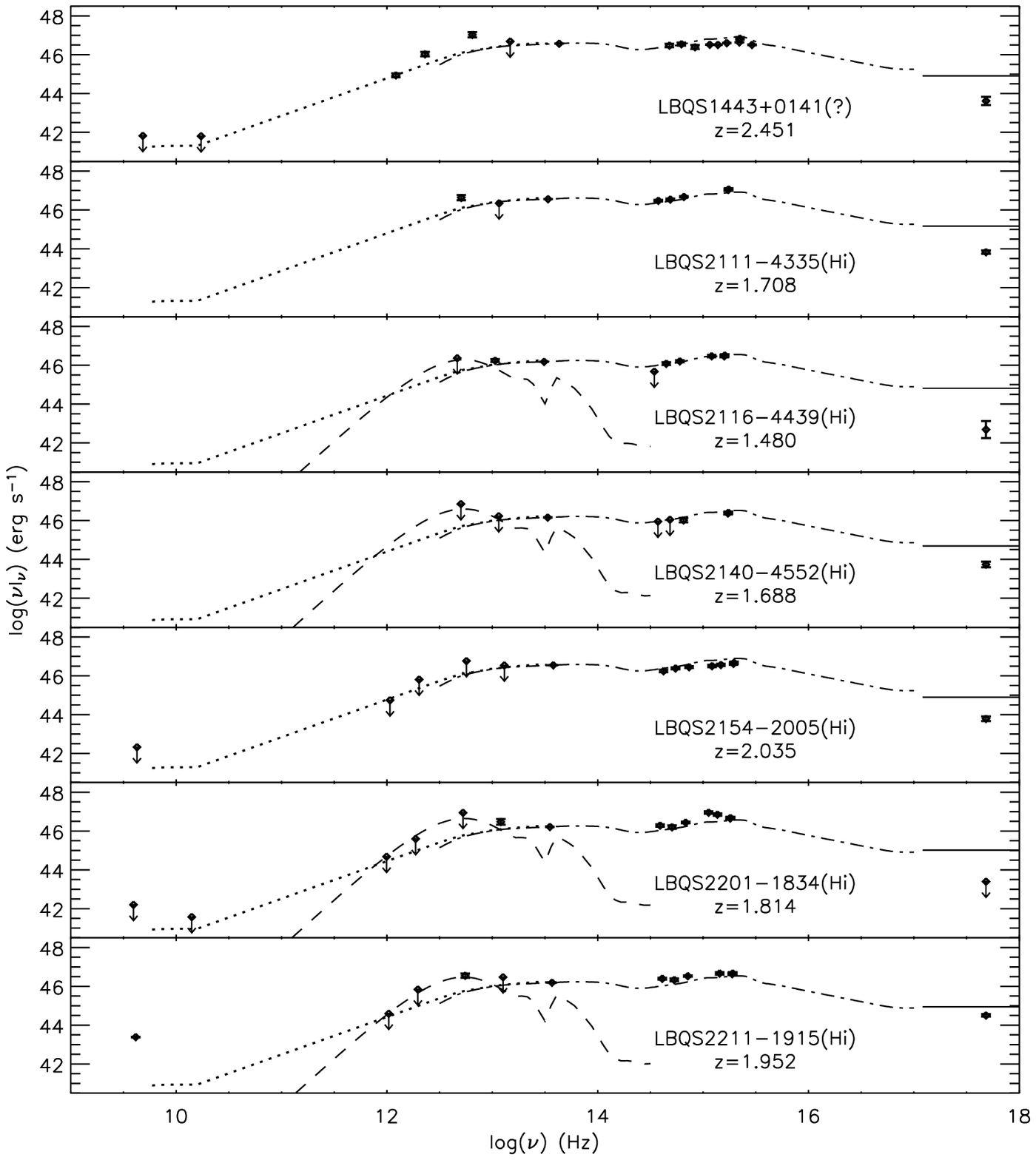}
\caption{
See Figure~\ref{fig:sed1} caption.}
\label{fig:sed5}
\end{figure*}

\begin{figure*}
\figurenum{1f}
\plotone{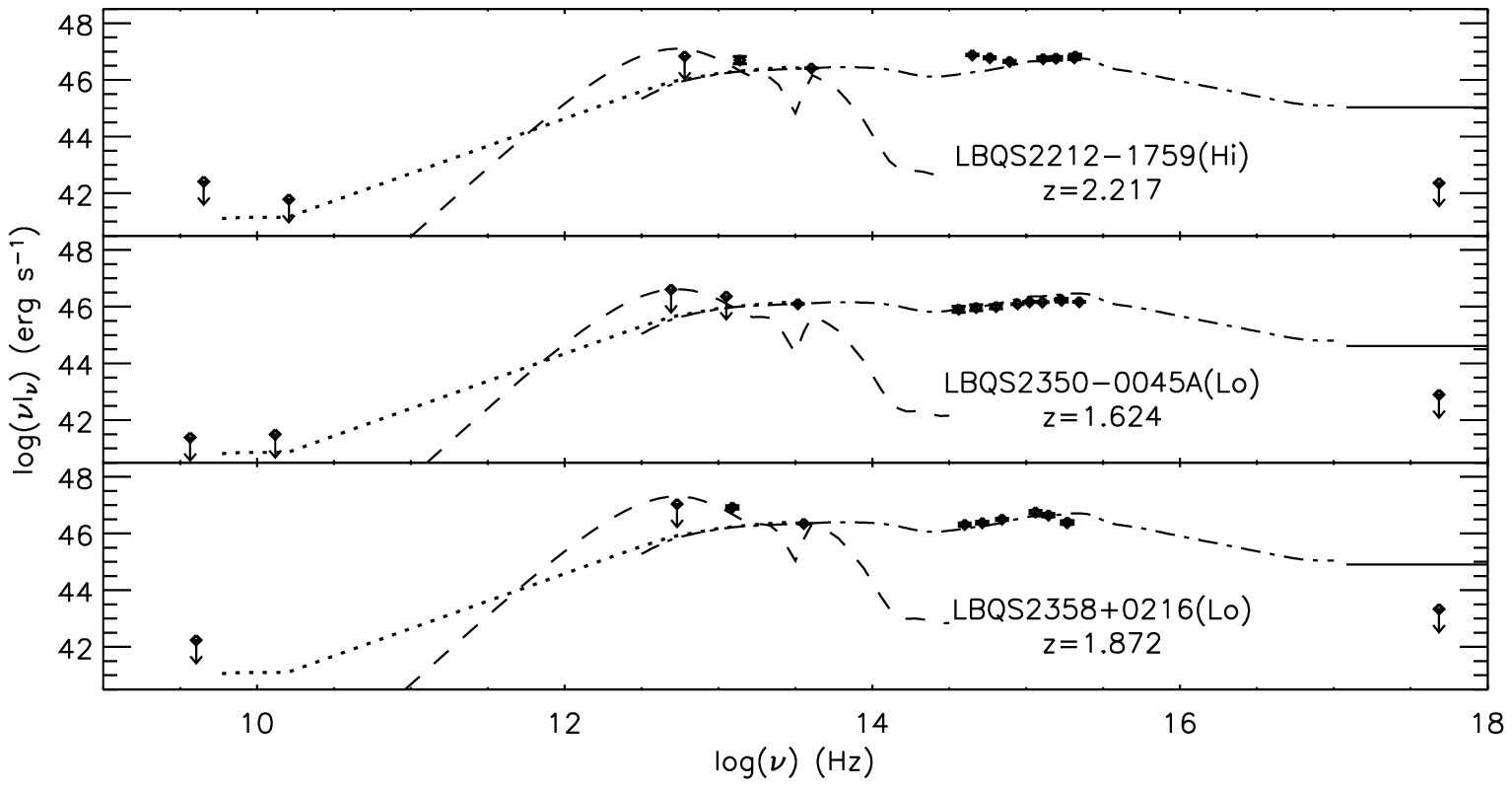}
\caption{
See Figure~\ref{fig:sed1} caption.}
\label{fig:sedlast}
\end{figure*}

We present the available multiwavelength photometry for each \balq\ in
Tables~\ref{tab:phot1} and \ref{tab:phot2}.  The SED of each quasar is
plotted in Figures~\ref{fig:sed1}--\ref{fig:sedlast} in rest-frame
units of $\log(\nu l_{\nu})$ (\lumin) vs. $\log(\nu)$ (Hz), where
$\log(\l_{\nu})$ includes the bandpass correction: $\log(f_{\nu}) -
\log(1+z) + \log(4\pi d_{\rm lum}^2) - 26$ ($f_{\nu}$ in mJy and
luminosity distance, $d_{\rm lum}$, in cm).  For reference, composite
SEDs normalized to the 24\micron\ MIPS data points are overlaid.  The
composite SEDs are the \citet{ElvisEtal1994} mean SED at
$\log(\nu)<13.57$ (8\micron) and the \citet{ric+06} luminous mean
quasar SED for $\log(\nu)=12.50$--17.00 (95\micron\ to 0.4~keV).  For
the $>0.5$~keV X-ray region, a spectral index of $\alpha_{\nu}=-1.0$
has been assumed \citep[e.g.,][]{GaBrChGa2002,PageEtal2005}, with the
normalization fixed by the best-fitting relation from
\citet[][eq. 1c]{steffen+06}: $\log(l_{\rm 2keV})=0.721\log(l_{\rm
2500}) + 4.531$.  This accounts for the empirical observation that the
most ultraviolet-luminous quasars emit relatively less power in
X-rays. From a qualitative inspection of the SEDs, the most obvious
discrepancy between the plotted composites and the data is the known
deficit at X-ray wavelengths when compared to the expected X-ray
power.

To investigate the nature of mid-infrared SEDs of \balqs, a comparison
sample is necessary.  Given the known SED changes of normal type~1
quasars with luminosity in the X-ray
\citep[e.g.,][]{AvTa1986,strateva+2005,steffen+06} and infrared
regimes \citep{ric+06,gall+06b}, identifying a comparison sample of
objects with comparable luminosity is clearly preferred. The largest
sample of quasar SEDs to date was presented by \citet{ric+06} who
incorporated the publicly available \spitzer\ MIPS+IRAC photometry as
of the Data Release 3 quasar catalog of \citet{dr3_qso}. All 259
objects were detected by \spitzer.  We draw our comparison from the
radio-quiet subset of this sample matched in monochromatic 5000\AA\
luminosity to the LBQS \balq\ sample ($\log(l_{\rm 5000
\AA})=31.21$--32.29); a total of 40 objects.  This luminous quasar
sample is referred to as the SDSS comparison sample hereafter.  While
ideally the comparison sample would be drawn from the LBQS itself to
mitigate possible selection effects, sensitive multiwavelength data as
required for the comparison do not exist elsewhere.  Luminous quasars
are quite rare, and so accumulating a sizable sample (of order tens) requires
multiwavelength, large area surveys or explicit targeting.

To compare directly the LBQS \balq\ sample with the SDSS comparison
sample, we compute three fiducial monochromatic luminosities at
rest-frame 8\micron, 5000\AA, and 2500\AA.  The first is measured by
normalizing the \citet{ric+06} luminous ($\log(L_{\rm
bol})>46.02$\lumin) quasar composite SED to the 24\micron\ MIPS
datapoint.  The second two are calculated from the best-fitting
normalization, $l_{\nu,0}$, and spectral index, \alphao, of a power-law model fit to the
available rest-frame 1200--5000\AA\ photometry: $l_\nu=l_{\rm
\nu,0}\nu^{\alpha_{\rm o}}$.  The lower frequency bound for this
ultraviolet-optical bandpass is chosen to avoid the optical spectral
region of $\lambda>5000$\AA\ ($\nu<10^{14.78}$~Hz) where host-galaxy
contamination can be significant \citep{VandenBerk2001}; at
$\lambda<1200$\AA ($\nu>10^{15.40}$~Hz), the intervening
Ly$\alpha$ forest affects the spectra considerably. For the \balqs,
the parameters, $\alpha_{\rm o}$, $\log(l_{\rm 8\mu{m}})$,
$\log(l_{\rm 5000 \AA})$, and $\log(l_{\rm 2500 \AA})$, are listed in
Table~\ref{tab:lums} along with $\log(l_{\rm 2keV})$ for those with
X-ray data.

For reference, two integrated infrared luminosities, $L_{\rm ir,QSO}$
(1mm to 2\micron) and $L_{\rm fir,QSO}$ (1mm to 20\micron), are
calculated using \leight\ and the composite SEDs of
\citet{ElvisEtal1994} (for $\lambda\ge40$\micron) and \altcite{ric+06} (for
$\lambda<40$\micron).  \citet{ElvisEtal1994} (which used \iras\ data)
has better coverage at longer wavelengths, though we point out that
neither composite is empirically constrained in the range from
$\log(\nu)=10.2$--12.5 (16~GHz to 95\micron).  Using the combined
composite SED, the infrared and far-infrared bolometric corrections
from $\nu$\leight\ to $L_{\rm ir,QSO}$ and $L_{\rm fir,QSO}$ are 2.99
and 0.84, respectively.  Given that the near and mid-infrared emission
is believed to be dominated by the quasar, these integrated infrared
luminosities (normalized at rest-frame $\sim8$\micron) may be
considered an estimate of the quasar's contribution to the infrared
SED.  Any additional power in this regime could therefore be attributed to
star-formation in the host galaxy.  

\subsection{Ultraviolet-Optical Comparison}

Spectroscopic comparisons of BAL and non-\balqs\ include the seminal
study by \citet{WeMoFoHe1991} and subsequent SDSS work by
\citet{reich+03b}.  While \citet{WeMoFoHe1991} concluded that the
underlying ultraviolet continuum and emission-line properties of BAL
quasars are consistent with those of normal quasars from their sample
of 42 \balqs, the larger (224) study of \citet{reich+03b} found that
\balqs\ are more specifically drawn from a parent sample of
intrinsically blue quasars.  A further caveat is the statistical
detection of excess dust extinction and reddening in the \balq\
population, most notably among Lo\balqs\ \citep{SpFo1992,trump+06}.

\begin{figure*}
\figurenum{2}
\plottwo{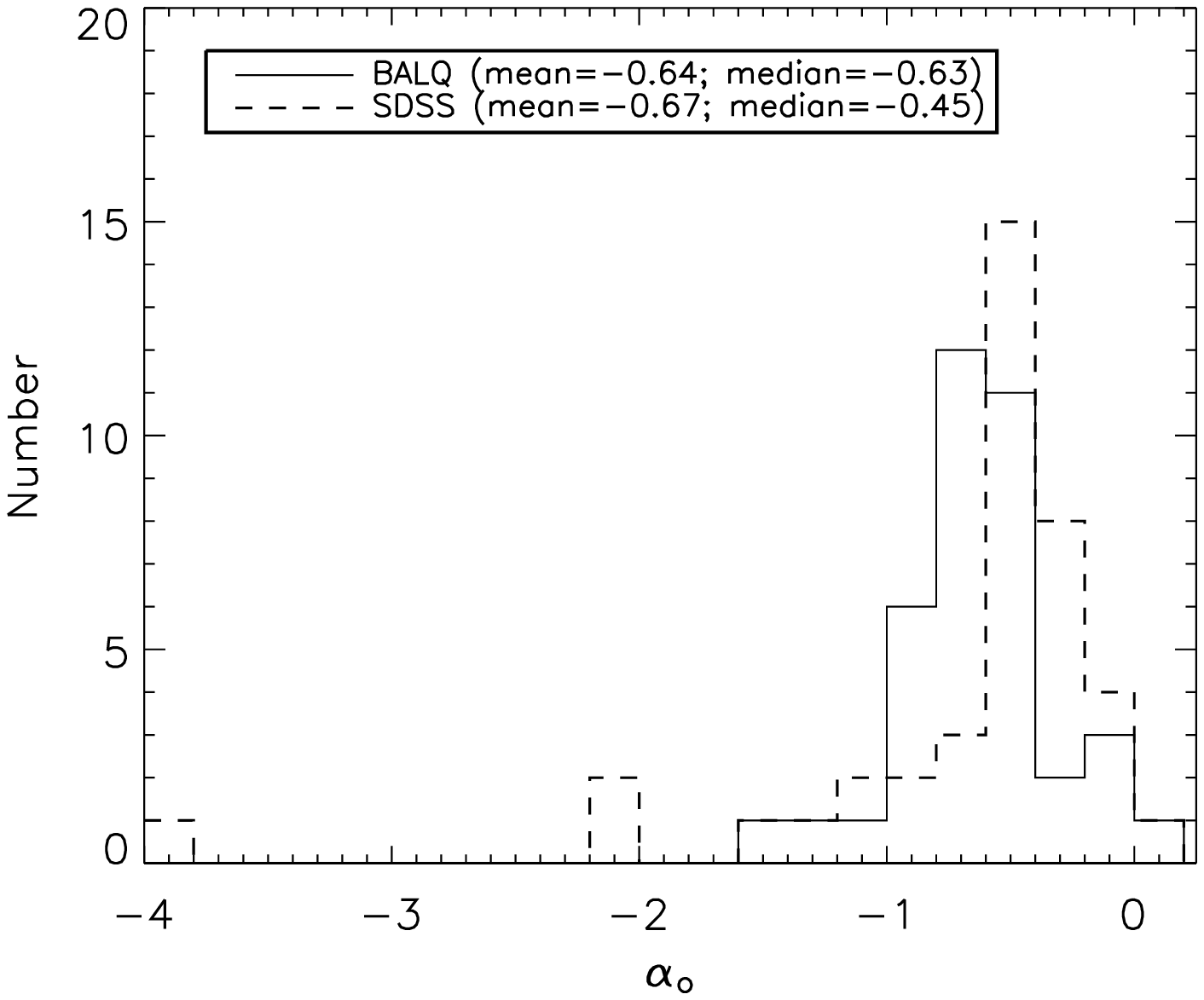}{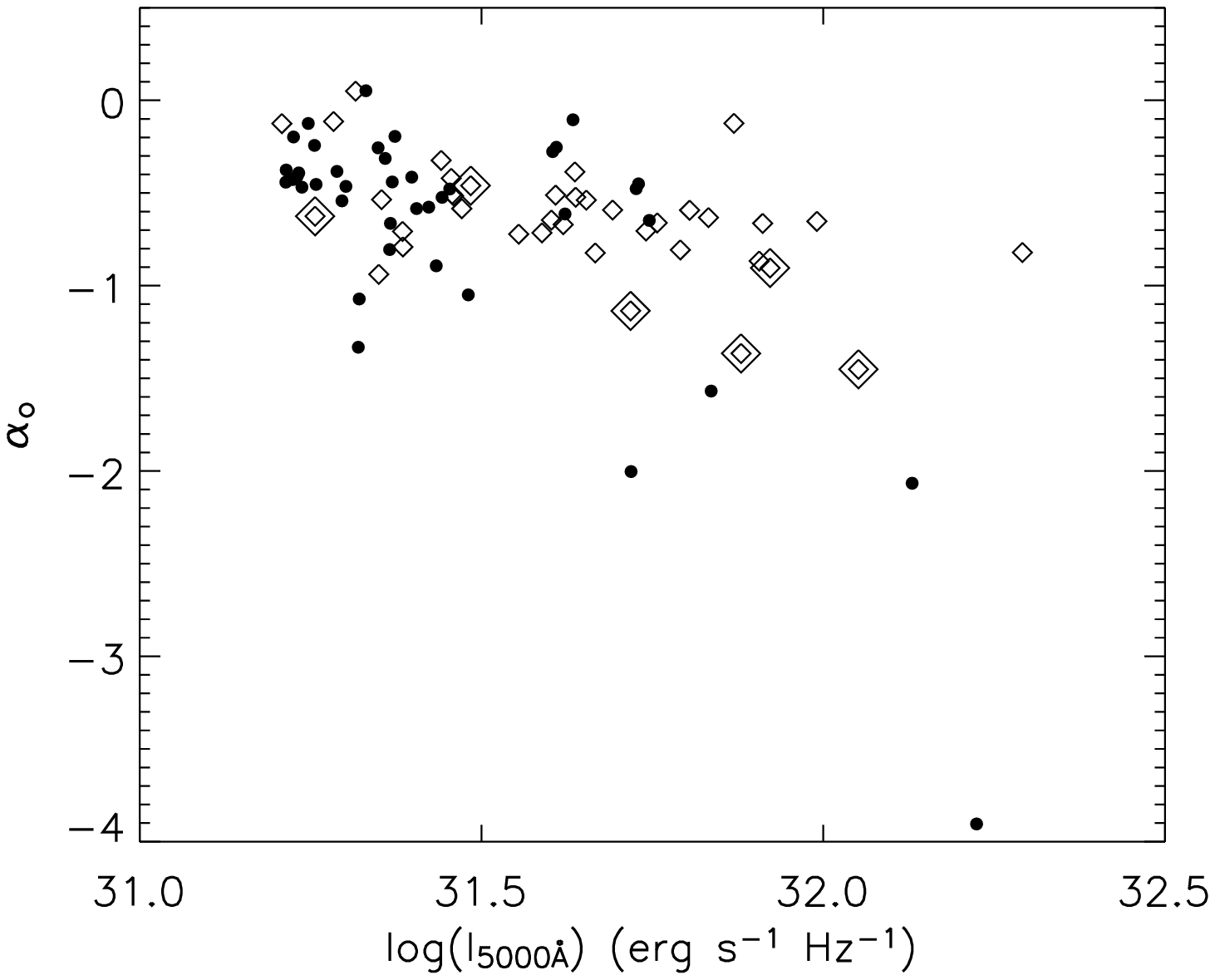}
\caption{
{\bf Left:}
Histograms of \alphao\ values for the BAL (solid) and SDSS
comparison (dashed) quasar samples. More negative values of \alphao\
indicate a redder ultraviolet-optical continuum. 
{\bf Right:}
A plot of \alphao\ vs. \lfive\ for the BAL (open diamonds) and SDSS
comparison (filled circles) samples.  Known Lo\balqs\ are indicated
with double diamonds. Both samples show a tendency for redder quasars
to be more luminous at 5000\AA; this is likely a selection effect.
}
\label{fig:opt}
\end{figure*}

In this study, we evaluate the ultraviolet-optical continua of BAL and
non-\balqs\ in a complementary fashion, using the rest-frame
1200--5000\AA\ photometry.  The inclusion of 2MASS data for the
\balqs\ and the SDSS $z$-band photometry enables an extension of
previous studies to the rest-frame optical.  First, we compare the
distribution of \alphao\ for both the LBQS BAL and SDSS comparison
quasar samples; histograms of the two distributions are shown in
Figure~\ref{fig:opt}.

To compare the two samples, we calculated the Student's $T$-statistic
of \alphao, $T=0.30$. This value indicates a probability of the null
hypothesis that the two samples have the same mean of 77\%.  However,
inspection of Figure~\ref{fig:opt} reveals that this statistic is
perhaps not capturing differences between the two samples: the \balqs\
seem to be slightly redder on average than normal quasars.  The median
values of --0.45 (SDSS) and --0.63 (\balq) reflect this apparent
difference.  To investigate if this is significant, we use the ASURV
implementation \citep{asurv_ref} of two non-parametric statistical
tests, the Wilcoxon and logrank, and obtain probabilities of the null
hypothesis (i.e., that the BAL and SDSS comparison samples have the
same distribution of \alphao) of 0.04 and 0.08, respectively.  These
results are not statistically significant evidence for a difference
between the samples.

To investigate further the \alphao\ distributions, we first examine
the effect of convolving a non-\balq\ composite spectrum from the SDSS
\citep{VandenBerk2001} with the SDSS filter functions in the redshift
range, $z=1.5$--2.9, and then measuring \alphao.  The strong
Ly$\alpha$ and \ion{C}{4} emission lines at the blue end of the
1200--5000\AA\ bandpass give a bluer \alphao\ of --0.34 (averaged over
the redshift range) than the value from fitting the continuum of the
SDSS composite spectrum, \mbox{$\alpha_{\rm o,spec}$=--0.44}
\citep{VandenBerk2001}.  Using the SDSS BAL and Lo\balq\ composite
spectra \citep{reich+03a,reich+03b} in the same manner, we obtain
\alphao=--0.67 and --1.28, respectively. The \balq\ composites give
redder values of \alphao\ because of both dust reddening and
extinction and the loss of flux on the blue end from \ion{C}{4} and
\ion{Si}{4} BALs.  Both of these effects are stronger in the Lo\balqs,
and the typical extinction measured in Hi and Lo\balq\ spectra is
$E(B-V)$=0.02 and 0.08, respectively \citep{reich+03b}.  This is
consistent with the trend in the data where the median \alphao\ values
are --0.59 and --0.90 for Hi and Lo\balqs, respectively.  We do not
expect the composite values to represent our data precisely, because
of the range of available photometric data as well as the inherent
dispersion in the ultraviolet-optical spectral properties of BAL and
non-\balqs.  

The right-hand panel of Figure~\ref{fig:opt} illustrates an
interesting artifact of optical quasar selection.  In the plots of
\alphao\ vs. \lfive, there is an apparent correlation between the two
whereby more optically luminous quasars are redder. This is a
statistically significant trend in both samples, the values of
Spearman's $\rho$ of $-0.498$ and $-0.452$ indicate probabilities that
the properties are not correlated of 0.0015 and 0.0034 for the BAL and
SDSS comparison quasar samples, respectively.  This is likely a
selection effect, as the identical test of \alphao\ vs. \ltwo\ yields
no evidence for a significant correlation with probabilities of 0.140
and 0.107 for the respective samples.  The monochromatic luminosity at
2500\AA\ is closer to the selection bandpasses of $B_J$ (for the LBQS)
and $i$ (for the SDSS) in this redshift range.  Of two quasars with
similar fluxes in the selection bandpass, the redder one will be more
luminous at longer wavelengths.

Overall, therefore, we find no compelling evidence for differences in
the ultraviolet-optical continua in BAL and non-\balqs\ beyond the
known tendency of BAL quasars to be mildly dust-reddened, in agreement
with previous optical spectroscopic studies.  As the optical quasar
selection criteria of the SDSS and the LBQS are different, there is
some concern that selection effects could be important.  Briefly, the
SDSS selection for radio-quiet quasars is based on an $i$-band flux
limit and broad-band photometric color selection, whereas the LBQS has
a $B_{J}$ flux limit and a complicated selection algorithm using
objective prism spectra.  LBQS quasar candidates were
identified via blue spectra and/or from features such as emission
lines and continuum breaks \citep[\S3.1.4;][]{lbqs_ref}.  Though the SDSS should
find all LBQS quasars, the converse is not necessarily true. (An
explicit comparison with the FIRST Bright Quasar Survey
[\altcite{fbqs_ref}] determined an incompleteness of 13$\pm4$\% for the
LBQS [\altcite{HeFoCh2001}]). Therefore, it is reassuring that only
four of the 40 SDSS comparison quasars have values of \alphao\ redder
than those of the \balqs. In subsequent analyses we examine whether
these four objects affect our results.

\subsection{Mid-Infrared Comparison}
\label{sec:mir}

\begin{figure*}
\figurenum{3}
\plottwo{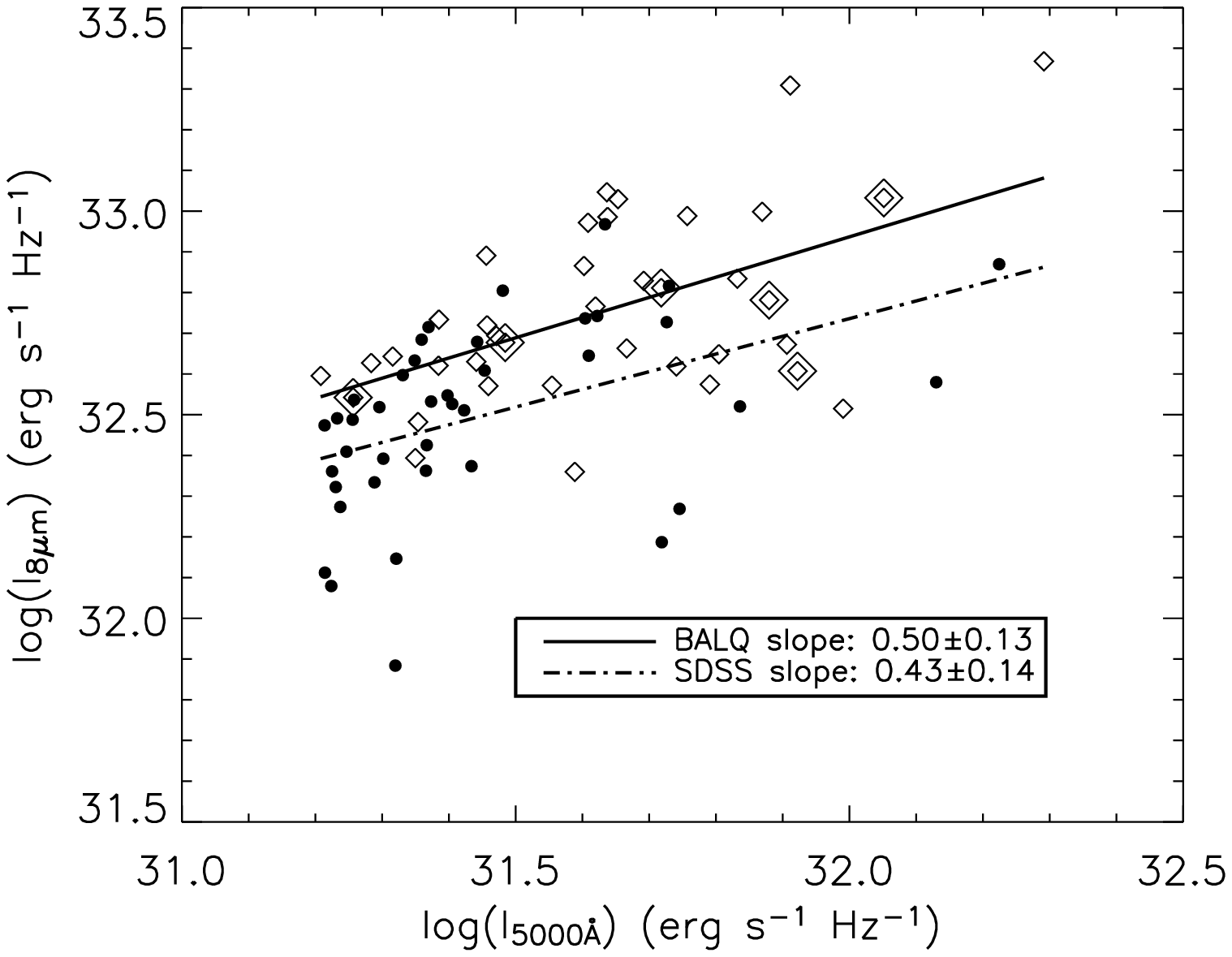}{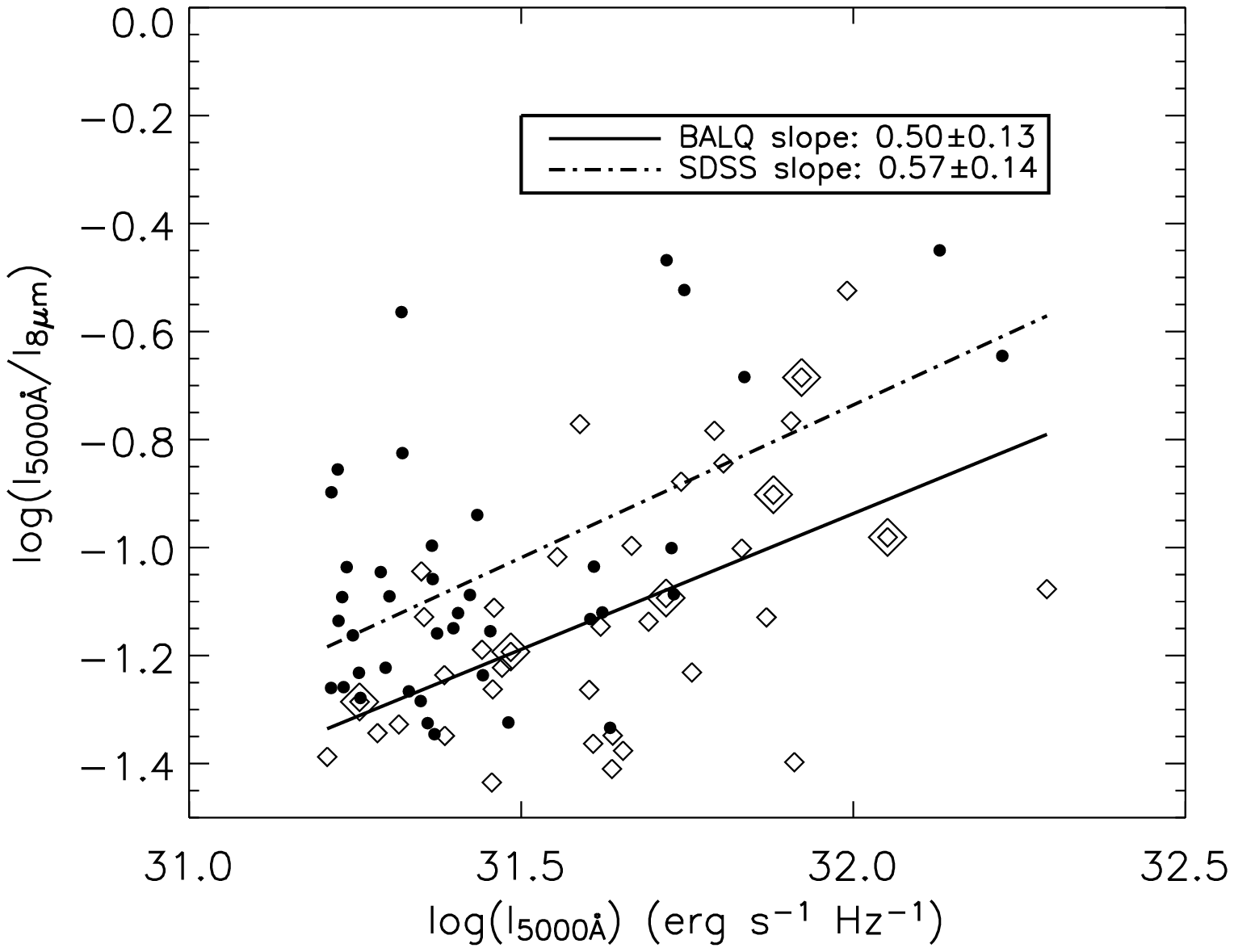}
\caption{
In both panels, the \balq\ (open diamonds; solid line) and SDSS comparison (filled
circles; dot-dashed line) samples are plotted.  Known Lo\balqs\ are
indicated with double diamonds. Best-fitting linear fits are drawn for
each sample, and slopes are labeled.
{\bf Left:}
Plot of \leight\ vs. \lfive.
The \balq\ and SDSS comparison samples have both consistent slopes and
normalizations ($17.1\pm4.1$ and $18.9\pm4.3$, respectively) in this
parameter space. Note that the 1$\sigma$ normalizations for both
samples cannot be plotted within the bounds of this figure.
{\bf Right:} 
Plot of \lfive/\leight\ vs. \lfive; this representation gives a better
idea of the relative luminosity of the optical versus the
mid-infrared.  Again, both slopes and normalizations are within 1$\sigma$
for the BAL and non-\balq\ samples, and the 1$\sigma$ normalizations
for both samples are outside of the figure bounds.
}
\label{fig:iropt}
\end{figure*}

From visual inspection of the mid-infrared SEDs presented in
Figure~1, the composite quasar SED
normalized at the 24\micron\ MIPS data point appears to predict quite
well the ultraviolet-optical photometry for most \balqs.  
As there is an inherent dispersion of quasar SEDs that is not captured in
the mean composite \citep{ElvisEtal1994,ric+06}, we quantify the
relationship between the mid-infrared and optical by comparing
\leight\ and \lfive\ for the BAL and SDSS comparison samples.
Specifically, we measure linear least-squares fits for log(\leight)
vs. log(\lfive) and log(\lfive/\leight) vs. log(\lfive). As
can be seen in Figure~\ref{fig:iropt}, both samples occupy the same
region of parameter space, and the normalizations and slopes are
consistent within 1$\sigma$. Note that by using \lfive\ rather than
\ltwo, we mitigate the effects of optical extinction; at 5000\AA, the
monochromatic luminosity is only reduced by $\sim7\%$ for
$E(B-V)=0.08$.

In addition to consistent fits with the data, we also measure the
Student's $T$-statistic of log(\lfive/\leight) for both samples:
$T=-1.39$. This value indicates a 17\%\ probability that the null
hypothesis (that the two samples have the same mean) is correct.
Inspection of histograms of the two distributions and similar medians
(\mbox{--1.09/--1.14} for SDSS/BAL) confirm this conclusion.  Finally,
the Wilcoxon and logrank tests give null hypothesis probabilities of
16\% and 17\%, respectively.  If the four SDSS comparison quasars with
values of \alphao\ redder than the \balq{s} are excluded, the
Student's $T$-test significance increases to 73\% that the samples
have the same mean, and the medians become almost identical: --1.13
versus --1.14.  Therefore, we find no evidence of a discrepancy in
mid-infrared relative to optical power for BAL and non-\balqs\ of
comparable luminosity.

\subsection{Far-Infrared Properties}
\label{sec:fir}

From inspection of Figure~1, the
far-infrared ($\lambda<20$\micron) properties of the \balqs\ are not
homogeneous.  Specifically, if one focuses on the quasars with 70,
160\micron\ and/or submm detections (rather than just upper limits),
there is a notable dispersion in relative and absolute far-infrared
power.  This is in contrast to the mid-infrared through optical
regimes, where the \balq\ SEDs much more closely follow the composite
SED. Though most of the 70 and 160\micron\ MIPS and SCUBA data points
are upper limits, several are sensitive enough to put useful limits on
far-infrared emission in excess of a composite quasar SED.  To
estimate the possible contribution of star formation, we subtract the
composite quasar SED normalized to the 24\micron\ MIPS point from the
far-infrared photometry.  An illustrative starburst SED 
from \citet{efstath+00} is then normalized to the excess emission and
integrated from 1mm to 20\micron\ to obtain $L_{\rm fir,SF}$.  The
starburst model, characterized by a burst age of 16~Myr and a visual
optical depth of 150, was generated for the prototype
far-infrared/submm-hyperluminous $z>4$ radio-quiet quasar
BR1202$-$0725. This object is sampled densely enough in the
far-infrared through submm region that, unlike the majority of high
redshift, submm-luminous quasars, well-constrained model fits to its
photometric data can be obtained \citep{PrMc2001}.  Ideally, one would
prefer to generate individual fits to the far-infrared data, however,
the data quality available here does not warrant it.

The $L_{\rm fir,SF}$ values are listed in Table~\ref{tab:lums}, and
the starburst models are overplotted in
Figure~1.  Upper limits are given when
there are no far-infrared detections or the data are consistent with a
quasar SED.  In the latter case, $L_{\rm fir,SF}$ was set to
$0.2L_{\rm fir,QSO}$ to take into account reasonable photometric
uncertainties -- the best-fitting values were significantly lower.

\begin{figure*}
\figurenum{4}
\plotone{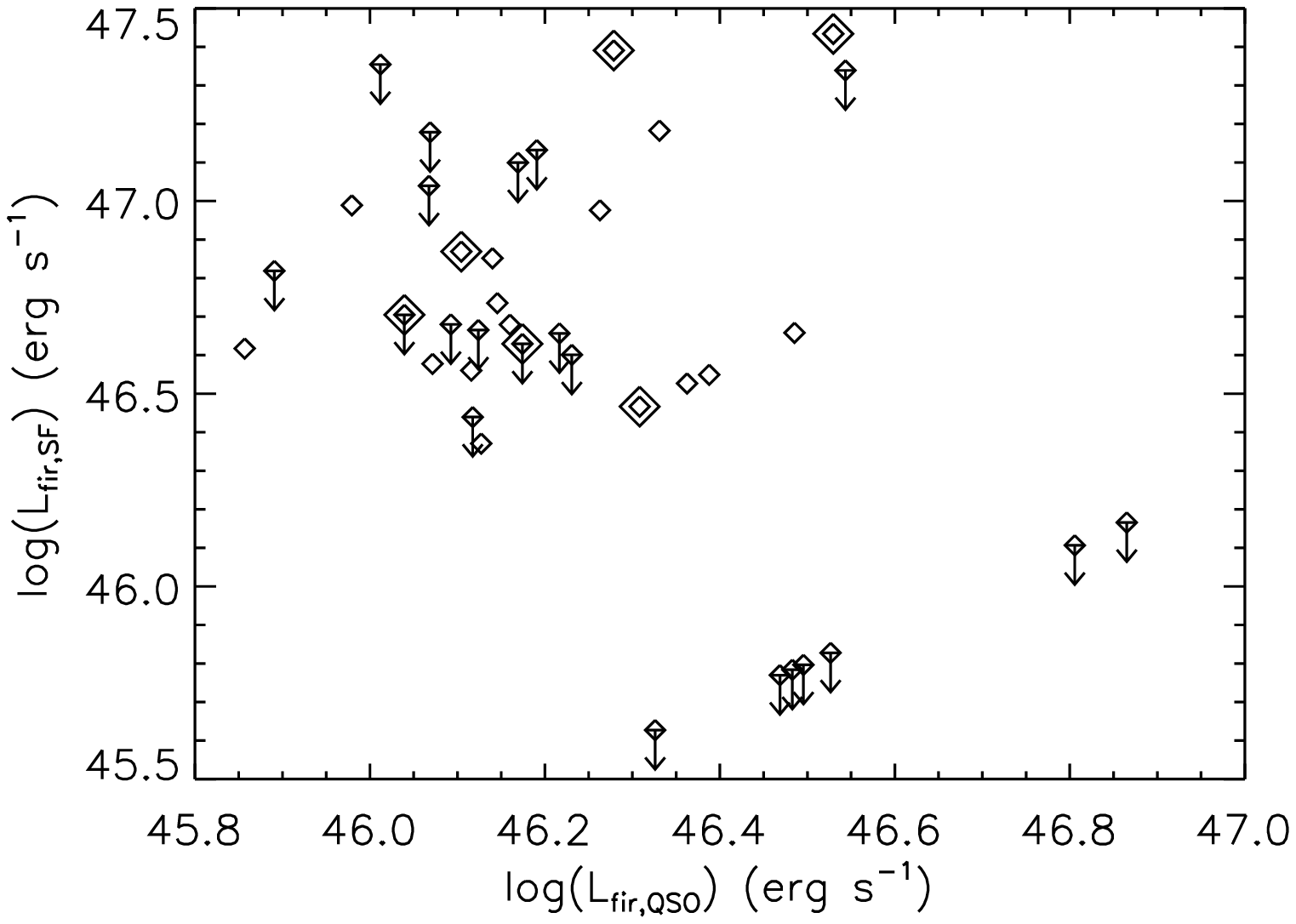}
\caption{
The far-infrared (1mm to 20\micron) integrated luminosity in star
formation, $L_{\rm fir,SF}$ vs. quasar far-infrared luminosity,
$L_{\rm fir,QSO}$. Double diamonds indicate Lo\balqs.  Upper limits in
the lower right of the plot were set to 20\%\ of $L_{\rm fir,QSO}$
when the data were consistent with the composite quasar SED.  Note
that the apparent gap in the distribution is from the 
disparate data available; all of the quasars with tight upper limits to $L_{\rm
fir,SF}$ have SCUBA data except for 2111$-$4335, which has a
constraining 70\micron\ upper limit.  Presumably with complete submm
coverage the distribution would more evenly fill this parameter space.}
\label{fig:fir}
\end{figure*}

In Figure~\ref{fig:fir}, the estimated contributions from star
formation are plotted against the quasar far-infrared power. Of 38
\balqs\ in the sample, 22 (58\%) have only upper limits to $L_{\rm
fir,SF}$, of these, seven (18\%) have far-infrared constraints
sufficient to determine that the quasar likely dominates their
far-infrared emission.  Based on local examples of a few Lo\balqs\
identified with ULIRGS, an anecdotal identification of Lo\balqs\ with
merger remnants and starburst activity had been suggested
\citep[e.g.][]{LoHuKlCu1988,CaSt2001}.  Though the two most luminous
starbursts are the Lo\balqs\ 1331$-$0108 and 2358$+$0216, the other
four are not obviously different from the Hi\balq\ population.
However, the preponderance of upper limits at far-infrared wavelengths
for the majority of Lo and HiBAL objects hampers our
ability to evaluate this topic further.

Of the 17 (45\%) \balqs\ with $L_{\rm fir,SF}$ detections, their
$L_{\rm fir,SF}$ values range from 1.4--13$L_{\rm fir,QSO}$ or
$10^{13-14}$\lsun, qualifying them as hyperluminous starbursts.  Using
the standard conversion\footnote{The uncertainty in this relation is
dominated primarily by the assumed initial mass function
(cf. \altcite{PrMc2001}).}  to star formation rates,
$SFR\approx10^{-10} L_{\rm fir,SF}/$\lsun\ (M$_{\odot}$~yr$^{-1}$),
gives $SFR=10^{3-4}$ M$_{\odot}$~yr$^{-1}$.  Given the upper limits,
even the objects without a detected far-infrared excess could host
substantial starbursts.  However, the uncertain origin of far-infrared
emission in quasars, from cold dust heated by star formation and/or
accretion power (see \altcite{Ro2000} and \altcite{haas+03} for
opposite views on this matter), means that a far-infrared `excess'
could be accretion powered.

\subsection{A Mid-Infrared-to-X-ray BAL Quasar Composite SED}
\label{sec:comps}

To synthesize the results presented in \S\ref{sec:seds} thus far, we
construct three composite SEDs, one for the SDSS comparison sample,
a second for the entire \balq\ sample of 38, and a third for the 25 known
Hi\balqs.  The composite from the SDSS comparison sample has been made
with photometric data (including IRAC) from the 36 quasars whose
values of \alphao\ are greater (bluer) than $-1.45$, the minimum for
the \balqs. With only six Lo\balqs, there were not enough data to
construct a composite SED for this type. The photometry from
24\micron\ through 1~keV is used; there are insufficient detections at
longer wavelengths to incorporate the far-infrared and submm.

As a first step, we normalize the SEDs for each object to \lfive\ (at
$\log(\nu)$=14.78); this is necessary to use the redshift spread
($z=1.5$--2.9) to fill out the multiwavelength wavelength coverage.
Next, we use a variable frequency window sized to contain at least
seven photometric data points (though a few bins at the high frequency
ends of available photometric coverage have fewer data points).
Within each bin, the median normalized luminosity is identified.  In
Figure~\ref{fig:comps}, we plot the normalized data for all \balqs, as
well as our SDSS and
\balq\ composites.  The
\citet{ric+06} non-\balq\ composite is overplotted for reference.  The
X-ray regime of the SDSS composite has been appropriately normalized
to match the median \mbox{$\log$(\ltwo)} (31.42\lumin) \balq\
luminosity as described in the beginning of \S\ref{sec:seds}.

From the \balq\ composite, the mid-infrared emission shows no evidence
for being brighter than expected based on the 5000\AA\ luminosity in
comparison with the SDSS composite. In the optical-ultraviolet region,
the structure of the composites follows the bright emission lines,
including H$\alpha$, H$\beta$, \ion{Mg}{2}, \ion{C}{4}, and
Ly$\alpha$.  The reddened continuum in comparison with the
\citet{ric+06} composite is also evident, as is very strong X-ray
absorption.  While the BAL and Hi\balq\ composites are not noticeably
different, the Lo\balq\ data cluster in the lower panel of
Figure~\ref{fig:comps}.  In particular, the ultraviolet-optical
photometry indicates stronger dust reddening and extinction and the
extreme X-ray weakness is pronounced.  The mid-infrared behavior bears
further investigation, as some of the 24\micron\ data appear to lie at
the faint end of the distribution. The composite SDSS comparison and
\balq\ SEDs are listed in Table~\ref{tab:comps}.

\section{Summary and Conclusions}

\begin{figure*}
\figurenum{5}
\plotone{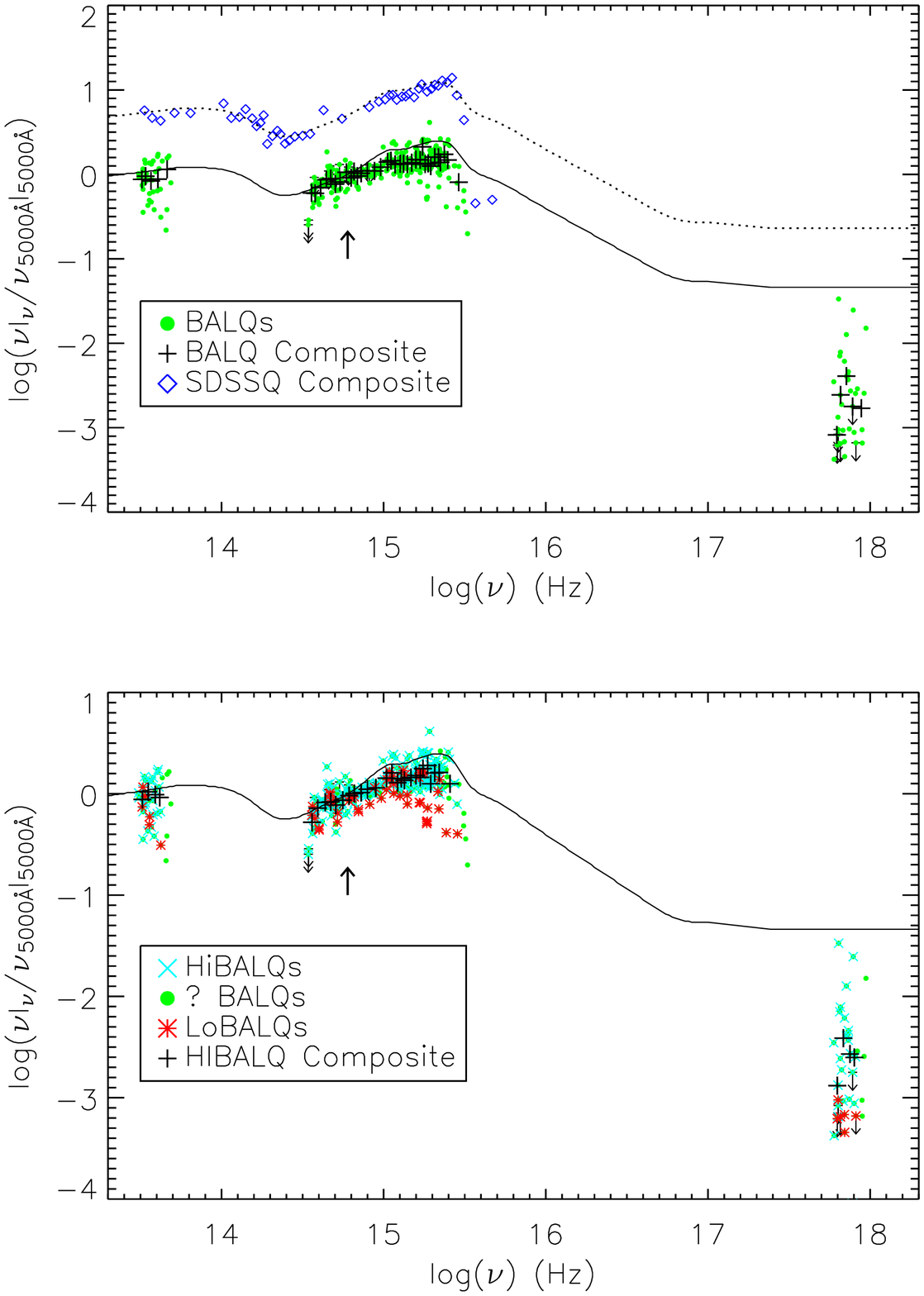}
\caption{ Mid-infrared-to-X-ray \balq\ composite SEDs.  The algorithm
  for constructing the composite SEDs is described in
  \S\ref{sec:comps}.  The data and the \citet{ric+06} SDSS composite
  SED (black curve) have been normalized to $\nu$\lfive.  For
  reference, the 5000\AA\ normalization point ($\log(\nu)=14.78$) is
  marked with an arrow, and 8 \microns\ is at $\log(\nu)=13.57$.  {\bf
  Top panel:} All \total\ \balqs\ (green filled circles) and their
  composite (black crosses); the SDSS comparison sample composite
  (blue diamonds) and its corresponding \citet{ric+06} SED (dotted
  curve) are offset vertically for clarity. {\bf Bottom panel:} The
  Hi\balq\ composite SED (black crosses) and the \balq\ data
  delineated by BAL type: Hi (cyan $\times$), unknown (green filled
  circles), and Lo (red $\ast$).  }
\label{fig:comps}
\end{figure*}

Using existing archives and literature as well as new \spitzer\ MIPS
observations, we present 38 \balq\ spectral energy distributions, the
largest sample of radio through X-ray data of \balqs\ to date.  We
compare this sample to a radio-quiet SDSS quasar sample matched in
5000\AA\ luminosity whose data were first presented by \citet{ric+06}.
In agreement with the conclusions of \citet{WeMoFoHe1991} and
\citet{reich+03b}, we find that luminous, optically selected BAL and non-\balqs\
have consistent ultraviolet-optical continua once dust reddening is
taken into account.  In addition, the relative mid-infrared and
optical power in both populations are indistinguishable.  As already
known, the most notable difference between BAL and non-\balq\ SEDs is
in the X-ray regime, where \balqs\ are weak X-ray emitters as a result
of intrinsic absorption \citep[e.g.,][]{gall+06}.  All of these
characteristics are evident in the composite \balq\ and Hi\balq\ SEDs
presented in Figure~\ref{fig:comps} and Table~\ref{tab:comps}.

Quasar mid-infrared emission is believed to arise on spatially large
($>1$~pc) scales by reprocessing of the ultraviolet-optical continuum
emission by a cold, dusty medium.  In the picture where the covering
fraction of a dusty BAL wind is large, one might expect \balqs\ to be
relatively brighter in the mid-infrared than non-\balqs\ with little
or no wind because of a larger emitting volume of dust.  For
reference, if all (i.e., assuming a 4$\pi$ covering fraction for the
dust) of the 1200--5000\AA\ continuum luminosity absorbed by dust with
the typical extinction found for Hi and Lo\balqs\ ($E(B-V)=0.023$ and
0.077, respectively; \altcite{reich+03b}) were reemitted in the
mid-infrared, Hi (Lo) \balqs\ should be 18\% (47\%) brighter than
typical quasars.  This picture is not upheld by our data.

An understanding that \balqs\ seen along equatorial lines of sight
should have preferentially fainter and redder ultraviolet-optical
continua than normal quasars, by the mechanism of accretion-disk limb
darkening \citep{KrVo1998} or orientation-dependent, gray attenuation
\citep{Goodrich1997}, leads to a similar expectation of mid-infrared
bright \balqs.  Again, our results do not support this.  However, the
characterization of disk-wind models as generating `equatorial'
outflows is perhaps taken too literally; the lines of sight closest to
the accretion disk are likely completely blocked by the hot dust that
obscures the inner accretion region in type~2 (narrow-line) quasars
and generates the mid-infrared emission.  Furthermore, radiatively
driven winds are radial with the inclination angle of the outflow
dependent on the launching radius: winds launched from closer to the
black hole have a smaller inclination angle with respect to the disk
normal \citep{PrStKa2000,everett05}.  Therefore, even in the disk-wind
paradigm where limb-darkening and orientation-dependent attenuation
are operating, it is unclear that the inclination angles of BAL and
non-\balqs\ would be sufficiently different to detect a mid-infrared
excess with the data in hand.

Based on the comparison of mid-infrared and optical properties of BAL
vs. non-\balqs, well-designed, wide-angle, mid-infrared quasar surveys
offer the most promise for capturing the `true' \balq\ fraction of
type~1 quasars.  Near-infrared surveys sampling the rest-frame
optical may also prove fruitful in the future, but are currently not
nearly sensitive enough. Optically selected surveys will inevitably be
somewhat incomplete because the ultraviolet BALs along with any dust
extinction conspire to make \balqs\ fainter than normal quasars of
comparable intrinsic luminosity.  Clearly, soft-X-ray surveys will
suffer from similar (though much more extreme) problems because of the
severe effects of intrinsic absorption.  At higher X-ray energies,
absorption becomes much less of a factor.  However, the potential of
hard ($>10$~keV) X-ray surveys to obtain similar completeness to
\spitzer\ will not be realized for many years -- $z\sim2$ quasars are
not within the flux limits of current facilities.  In any case, there
is evidence for Compton-thick (\nh$>1.5\times10^{24}$\cmsq) X-ray
absorption in some \balqs\ that will affect even $>10$~keV X-rays
\citep[e.g.,][]{gall+06}.

Given the dearth of far-infrared coverage (particularly detections) of
non-\balqs\ of similar redshifts and luminosities to this sample, it
is unclear how the far-infrared properties of \balqs\ would fit within
a matched type~1 population.  At lower redshifts and luminosities, the
contribution of star formation to the far-infrared power of quasars is
quite variable, and many (non-BAL) quasars show evidence for quite
luminous starbursts \citep{fritz+06,schweitz+06}.  Even though
\spitzer\ is significantly more sensitive than \iras, the confusion
limit for MIPS is typically brighter than the expected 70 and
160\micron\ fluxes for luminous, $z\sim2$ quasars.  This is
unfortunate, given that the far-infrared is one spectral region likely
to show conclusive evidence of a star-forming host galaxy.  We draw
attention, however, to the value of submm data to set important
constraints on excess far-infrared emission; six of the seven \balqs\
with the tightest upper limits to $L_{\rm fir,SF}$ have sensitive
SCUBA data.  Thus the advent of the next generation submm and
far-infrared detectors will shed new light on this issue.

Though future studies with IRAC and IRS on \spitzer\ may reveal subtle
differences in BAL vs. non-\balq\ 1--8\micron\ continuum properties,
the gross, intrinsic properties of the SEDs of the two populations are
fully consistent. Two regimes where they do differ, in the tendency to
exhibit optical dust reddening and strong X-ray absorption, can both
be accounted for with orientation-dependent absorption in a wind; the
ultraviolet-optical and X-ray continua are not intrinsically
different. To date therefore, the disk-wind paradigm where all quasars
have a BAL region, and the detection of such is merely an accident of
orientation, has not been overturned.

\acknowledgements

We thank Paul Hewett, Kirk Korista, and Simon Morris for generously
providing access to electronic spectra.  Comments from Gordon
Richards, Pat Hall, and the anonymous referee improved this paper. We
acknowledge the following archives: the Sloan Digital Sky Survey
(http://www.sdss.org), 2MASS (http://www.ipac.caltech.edu/2mass),
FIRST (http://sundog.stsci.edu), and NVSS
(http://www.cv.nrao.edu/nvss). Support for S. C. G. was provided by
NASA through the Spitzer Fellowship Program, under award
1256317. W. N. B. acknowledges NASA LTSA grant NAG5--13035.



\begin{thebibliography}{62}\setlength{\itemsep}{0mm}
\expandafter\ifx\csname natexlab\endcsname\relax\def\natexlab#1{#1}\fi
\expandafter\ifx\csname url\endcsname\relax
  \def\url#1{{\tt #1}}\fi
\expandafter\ifx\csname urlprefix\endcsname\relax\def\urlprefix{URL }\fi
\providecommand{\eprint}[2][]{\url{#2}}

\bibitem[\protect\astroncite{{Avni} \& {Tananbaum}}{1986}]{AvTa1986}
{Avni}, Y. \& {Tananbaum}, H. 1986, \apj, 305, 83

\bibitem[\protect\astroncite{{Barvainis}}{1987}]{barvainis87}
{Barvainis}, R. 1987, \apj, 320, 537

\bibitem[\protect\astroncite{{Becker} et~al.}{2000}]{BeckerEtal2000}
{Becker}, R.~H., {White}, R.~L., {Gregg}, M.~D., {Brotherton}, M.~S.,
  {Laurent-Muehleisen}, S.~A., \& {Arav}, N. 2000, \apj, 538, 72

\bibitem[\protect\astroncite{{Brotherton} et~al.}{2006}]{brotherton+06}
{Brotherton}, M.~S., {de Breuck}, C., \& {Schaefer}, J.~J. 2006, \mnras, L90

\bibitem[\protect\astroncite{{Canalizo} \& {Stockton}}{2001}]{CaSt2001}
{Canalizo}, G. \& {Stockton}, A. 2001, \apj, 555, 719

\bibitem[\protect\astroncite{{Clavel} et~al.}{2006}]{clavel+06}
{Clavel}, J., {Schartel}, N., \& {Tomas}, L. 2006, \aap, 439

\bibitem[\protect\astroncite{{Condon} et~al.}{1998}]{nvss_ref}
{Condon}, J.~J., {Cotton}, W.~D., {Greisen}, E.~W., {Yin}, Q.~F., {Perley},
  R.~A., {Taylor}, G.~B., \& {Broderick}, J.~J. 1998, \aj, 115, 1693

\bibitem[\protect\astroncite{{Cristiani} \& {Vio}}{1990}]{CrVi1990}
{Cristiani}, S. \& {Vio}, R. 1990, \aap, 227, 385

\bibitem[\protect\astroncite{{Edelson} \& {Malkan}}{1986}]{EdMa1986}
{Edelson}, R.~A. \& {Malkan}, M.~A. 1986, \apj, 308, 59

\bibitem[\protect\astroncite{{Efstathiou} et~al.}{2000}]{efstath+00}
{Efstathiou}, A., {Rowan-Robinson}, M., \& {Siebenmorgen}, R. 2000, \mnras,
  313, 734

\bibitem[\protect\astroncite{{Elvis} et~al.}{1994}]{ElvisEtal1994}
{Elvis}, M., et~al. 1994, \apjs, 95, 1

\bibitem[\protect\astroncite{{Everett}}{2005}]{everett05}
{Everett}, J.~E. 2005, \apj, 631, 689

\bibitem[\protect\astroncite{{Fritz} et~al.}{2006}]{fritz+06}
{Fritz}, J., {Franceschini}, A., \& {Hatziminaoglou}, E. 2006, \mnras, 366, 767

\bibitem[\protect\astroncite{{Gallagher} et~al.}{2002}]{GaBrChGa2002}
{Gallagher}, S.~C., {Brandt}, W.~N., {Chartas}, G., \& {Garmire}, G.~P. 2002,
  \apj, 567, 37

\bibitem[\protect\astroncite{{Gallagher} et~al.}{2006}]{gall+06}
{Gallagher}, S.~C., {Brandt}, W.~N., {Chartas}, G., {Priddey}, R., {Garmire},
  G.~P., \& {Sambruna}, R.~M. 2006, \apj, 644, 709

\bibitem[\protect\astroncite{{Gallagher} et~al.}{2007}]{gall+06b}
{Gallagher}, S.~C., {Richards}, G.~T., {Lacy}, M., Hines, D.~C., Elitzur, M.,
  \& Storrie-Lombardi, L.~J. 2007, \apj, in press, astro-ph/0702272

\bibitem[\protect\astroncite{{Goodrich}}{1997}]{Goodrich1997}
{Goodrich}, R.~W. 1997, \apj, 474, 606

\bibitem[\protect\astroncite{{Granato} et~al.}{2004}]{GranatoEtal2004}
{Granato}, G.~L., {De Zotti}, G., {Silva}, L., {Bressan}, .~A., \& {Danese}, L.
  2004, \apj, 600, 580

\bibitem[\protect\astroncite{{Green} \& {Mathur}}{1996}]{GrMa1996}
{Green}, P.~J. \& {Mathur}, S. 1996, \apj, 462, 637

\bibitem[\protect\astroncite{{Gregg} et~al.}{2002}]{gregg+02}
{Gregg}, M.~D., {Becker}, R.~H., {White}, R.~L., {Richards}, G.~T., {Chaffee},
  F.~H., \& {Fan}, X. 2002, \apjl, 573, L85

\bibitem[\protect\astroncite{{Haas} et~al.}{2003}]{haas+03}
{Haas}, M., et~al. 2003, \aap, 402, 87

\bibitem[\protect\astroncite{{Hamann} et~al.}{1993}]{HaKoMo1993}
{Hamann}, F., {Korista}, K.~T., \& {Morris}, S.~L. 1993, \apj, 415, 541

\bibitem[\protect\astroncite{{Hewett} \& {Foltz}}{2003}]{HewFol2003}
{Hewett}, P.~C. \& {Foltz}, C.~B. 2003, \aj, 125, 1784

\bibitem[\protect\astroncite{{Hewett} et~al.}{1995}]{lbqs_ref}
{Hewett}, P.~C., {Foltz}, C.~B., \& {Chaffee}, F.~H. 1995, \aj, 109, 1498

\bibitem[\protect\astroncite{{Hewett} et~al.}{2001}]{HeFoCh2001}
--- 2001, \aj, 122, 518

\bibitem[\protect\astroncite{{Hopkins} et~al.}{2005}]{hopkins+05}
{Hopkins}, P.~F., {Hernquist}, L., {Martini}, P., {Cox}, T.~J., {Robertson},
  B., {Di Matteo}, T., \& {Springel}, V. 2005, \apjl, 625, L71

\bibitem[\protect\astroncite{{Korista} et~al.}{1993}]{KoVoMoWe1993}
{Korista}, K.~T., {Voit}, G.~M., {Morris}, S.~L., \& {Weymann}, R.~J. 1993,
  \apjs, 88, 357

\bibitem[\protect\astroncite{{Krolik} \& {Voit}}{1998}]{KrVo1998}
{Krolik}, J.~H. \& {Voit}, G.~M. 1998, \apjl, 497, L5

\bibitem[\protect\astroncite{{La~Valley} et~al.}{1992}]{asurv_ref}
{La~Valley}, M., {Isobe}, T., \& {Feigelson}, E. 1992, in ASP Conf. Ser. 25:
  Astronomical Data Analysis Software and Systems I, eds. D.~M. Worrall,
  C.~Biemesderfer, \& J.~Barnes, vol.~1, 245

\bibitem[\protect\astroncite{{Lamy} \& {Hutsem{\' e}kers}}{2004}]{LaHu2004}
{Lamy}, H. \& {Hutsem{\' e}kers}, D. 2004, \aap, 427, 107

\bibitem[\protect\astroncite{{Lewis} et~al.}{2003}]{lewis+03}
{Lewis}, G.~F., {Chapman}, S.~C., \& {Kuncic}, Z. 2003, \apjl, 596, L35

\bibitem[\protect\astroncite{{Low} et~al.}{1988}]{LoHuKlCu1988}
{Low}, F.~J., {Huchra}, J.~P., {Kleinmann}, S.~G., \& {Cutri}, R.~M. 1988,
  \apjl, 327, L41

\bibitem[\protect\astroncite{{Maddox} \& {Hewett}}{2006}]{bj_ref}
{Maddox}, N. \& {Hewett}, P.~C. 2006, \mnras, 367, 717

\bibitem[\protect\astroncite{{Murray} et~al.}{1995}]{MuChGrVo1995}
{Murray}, N., {Chiang}, J., {Grossman}, S.~A., \& {Voit}, G.~M. 1995, \apj,
  451, 498

\bibitem[\protect\astroncite{{Ogle} et~al.}{1999}]{OgCoMiTr1999}
{Ogle}, P.~M., {Cohen}, M.~H., {Miller}, J.~S., {Tran}, H.~D., {Goodrich},
  R.~W., \& {Martel}, A.~R. 1999, \apjs, 125, 1

\bibitem[\protect\astroncite{{Page} et~al.}{2005}]{PageEtal2005}
{Page}, K.~L., {Reeves}, J.~N., {O'Brien}, P.~T., \& {Turner}, M.~J.~L. 2005,
  \mnras, 898

\bibitem[\protect\astroncite{{Priddey} et~al.}{2007}]{priddey+06}
{Priddey}, R.~S., {Gallagher}, S.~C., {Isaak}, K.~G., Sharp, R.~G., {McMahon},
  R.~G., \& Butner, H.~M. 2007, \mnras, 374, 867

\bibitem[\protect\astroncite{{Priddey} \& {McMahon}}{2001}]{PrMc2001}
{Priddey}, R.~S. \& {McMahon}, R.~G. 2001, \mnras, 324, L17

\bibitem[\protect\astroncite{{Proga} et~al.}{2000}]{PrStKa2000}
{Proga}, D., {Stone}, J.~M., \& {Kallman}, T.~R. 2000, \apj, 543, 686

\bibitem[\protect\astroncite{{Reichard} et~al.}{2003{\natexlab{a}}}]{reich+03a}
{Reichard}, T.~A., et~al. 2003{\natexlab{a}}, \aj, 125, 1711

\bibitem[\protect\astroncite{{Reichard} et~al.}{2003{\natexlab{b}}}]{reich+03b}
--- 2003{\natexlab{b}}, \aj, 126, 2594

\bibitem[\protect\astroncite{{Richards} et~al.}{2002}]{ric+02}
{Richards}, G.~T., {Vanden Berk}, D.~E., {Reichard}, T.~A., {Hall}, P.~B.,
  {Schneider}, D.~P., {SubbaRao}, M., {Thakar}, A.~R., \& {York}, D.~G. 2002,
  \aj, 124, 1

\bibitem[\protect\astroncite{{Richards} et~al.}{2006}]{ric+06}
{Richards}, G.~T., et~al. 2006, \apjs, in press, \eprint{(astro-ph/0601558)}

\bibitem[\protect\astroncite{{Rowan-Robinson}}{2000}]{Ro2000}
{Rowan-Robinson}, M. 2000, \mnras, 316, 885

\bibitem[\protect\astroncite{{Sanders} et~al.}{1988}]{SaEtal1988}
{Sanders}, D.~B., {Soifer}, B.~T., {Elias}, J.~H., {Neugebauer}, G., \&
  {Matthews}, K. 1988, \apjl, 328, L35

\bibitem[\protect\astroncite{{Scannapieco} \& {Oh}}{2004}]{ScOh2004}
{Scannapieco}, E. \& {Oh}, S.~P. 2004, \apj, 608, 62

\bibitem[\protect\astroncite{{Schneider} et~al.}{2005}]{dr3_qso}
{Schneider}, D.~P., et~al. 2005, \aj, 130, 367

\bibitem[\protect\astroncite{{Schweitzer} et~al.}{2006}]{schweitz+06}
{Schweitzer}, M., et~al. 2006, \apj, in press, \eprint{(astro-ph/0606158)}

\bibitem[\protect\astroncite{{Skrutskie} et~al.}{2006}]{2mass_ref}
{Skrutskie}, M.~F., et~al. 2006, \aj, 131, 1163

\bibitem[\protect\astroncite{{Spergel} et~al.}{2003}]{sperg03}
{Spergel}, D.~N., et~al. 2003, \apjs, 148, 175

\bibitem[\protect\astroncite{{Spergel} et~al.}{2007}]{sperg06}
--- 2007, \apj, in press, \eprint{(astro-ph/0603449)}

\bibitem[\protect\astroncite{{Sprayberry} \& {Foltz}}{1992}]{SpFo1992}
{Sprayberry}, D. \& {Foltz}, C.~B. 1992, \apj, 390, 39

\bibitem[\protect\astroncite{{Springel} et~al.}{2005}]{SpDiHe2005}
{Springel}, V., {Di Matteo}, T., \& {Hernquist}, L. 2005, \mnras, 361, 776

\bibitem[\protect\astroncite{{Steffen} et~al.}{2006}]{steffen+06}
{Steffen}, A.~T., {Strateva}, I., {Brandt}, W.~N., {Alexander}, D.~M.,
  {Koekemoer}, A.~M., {Lehmer}, B.~D., {Schneider}, D.~P., \& {Vignali}, C.
  2006, \aj, 131, 2826

\bibitem[\protect\astroncite{{Stocke} et~al.}{1992}]{stocke+92}
{Stocke}, J.~T., {Morris}, S.~L., {Weymann}, R.~J., \& {Foltz}, C.~B. 1992,
  \apj, 396, 487

\bibitem[\protect\astroncite{{Strateva} et~al.}{2005}]{strateva+2005}
{Strateva}, I.~V., {Brandt}, W.~N., {Schneider}, D.~P., {Vanden Berk}, D.~G.,
  \& {Vignali}, C. 2005, \aj, 130, 387

\bibitem[\protect\astroncite{{Trump} et~al.}{2006}]{trump+06}
{Trump}, J.~R., et~al. 2006, \apjs, 165, 1

\bibitem[\protect\astroncite{{Vanden Berk} et~al.}{2001}]{VandenBerk2001}
{Vanden Berk}, D.~E., et~al. 2001, \aj, 122, 549

\bibitem[\protect\astroncite{{Weymann} et~al.}{1991}]{WeMoFoHe1991}
{Weymann}, R.~J., {Morris}, S.~L., {Foltz}, C.~B., \& {Hewett}, P.~C. 1991,
  \apj, 373, 23

\bibitem[\protect\astroncite{{White} et~al.}{1997}]{first_ref}
{White}, R.~L., {Becker}, R.~H., {Helfand}, D.~J., \& {Gregg}, M.~D. 1997,
  \apj, 475, 479

\bibitem[\protect\astroncite{{White} et~al.}{1997}]{fbqs_ref}
{White}, R.~L., et~al. 2000,  \apjs, 126, 133

\bibitem[\protect\astroncite{{Willott} et~al.}{2003}]{willott+03}
{Willott}, C.~J., {Rawlings}, S., \& {Grimes}, J.~A. 2003, \apj, 598, 909

\bibitem[\protect\astroncite{{York} et~al.}{2000}]{sdss_ref}
{York}, D.~G., et~al. 2000, \aj, 120, 1579

\bibitem[\protect\astroncite{{Zhou} et~al.}{2006}]{zhou+06}
{Zhou}, H., {Wang}, T., {Wang}, H., {Wang}, J., {Yuan}, W., \& {Lu}, Y. 2006,
  \apj, 639, 716

\end{thebibliography}

\clearpage
\pagestyle{empty}
\begin{deluxetable}{lccrrcccccr}
\tabletypesize{\small}
\tablewidth{0pt}
\tablecaption{\balq\ Sample Properties
\label{tab:lums}
}
\tablehead{
\colhead{Name} &
\colhead{} &
\colhead{BAL} &
\colhead{} &
\multicolumn{4}{c}{$\log(l_{\nu})$ (\lumin\ Hz$^{-1}$)} &
\multicolumn{3}{c}{$\log(L)$ (erg \persec)} \\
\colhead{(LBQS B)} &
\colhead{$z$\tablenotemark{a}} &
\colhead{Type\tablenotemark{b}} &
\colhead{$\alpha_{\rm o}$\tablenotemark{c}} &
\colhead{2 keV\tablenotemark{d}} &
\colhead{2500\AA\tablenotemark{e}} &
\colhead{5000\AA\tablenotemark{e}} &
\colhead{8 \micron\tablenotemark{f}} &
\colhead{IR, QSO\tablenotemark{g}} &
\colhead{FIR, QSO\tablenotemark{g}} &
\colhead{FIR, SF\tablenotemark{h}} 
}
\startdata
0004$+$0147 &    1.710 &  Lo & $ -0.46\pm 0.19$ & $< 25.25$ &  31.35 &  31.48 &  32.68 &  46.73 &  46.17 & $< 46.63 $\\
0009$+$0219 &    2.642 &   ? & $ -0.65\pm 0.10$ & $\cdots$  &  31.79 &  31.99 &  32.51 &  46.56 &  46.01 & $< 47.35 $\\
0019$+$0107 &    2.130 &  Hi & $ -0.66\pm 0.07$    &  26.03 &  31.56 &  31.76 &  32.99 &  47.04 &  46.49   &    46.66\\
0021$-$0213 &    2.293 &  Hi & $ -0.65\pm 0.07$    &  25.27 &  31.41 &  31.60 &  32.87 &  46.91 &  46.36   &    46.53\\
0025$-$0151 &    2.076 &  Hi & $ -0.81\pm 0.11$    &  25.64 &  31.55 &  31.79 &  32.57 &  46.62 &  46.07   &    46.58\\
0029$+$0017 &    2.253 &  Hi & $ -0.42\pm 0.09$    &  26.85 &  31.33 &  31.46 &  32.89 &  46.94 &  46.39   &    46.55\\
0051$-$0019 &    1.713 &  Hi & $ -0.79\pm 0.15$    &  26.25 &  31.15 &  31.38 &  32.73 &  46.78 &  46.23 & $< 46.60 $\\
0054$+$0200 &    1.872 &  Hi & $ -0.11\pm 0.27$    &  25.95 &  31.25 &  31.28 &  32.63 &  46.68 &  46.12 & $< 46.67 $\\
0059$-$2735 &    1.593 &  Lo & $ -1.14\pm 0.16$ & $< 25.49$ &  31.38 &  31.72 &  32.81 &  46.86 &  46.31   &    46.47\\
0106$-$0113 &    1.668 &  Hi & $ -0.51\pm 0.07$ & $\cdots$  &  31.30 &  31.46 &  32.57 &  46.62 &  46.07 & $< 47.04 $\\
0109$-$0128 &    1.758 &  Hi & $ -0.58\pm 0.04$    &  25.60 &  31.30 &  31.47 &  32.69 &  46.74 &  46.19 & $< 47.13 $\\
1029$-$0125 &    2.029 &  Hi & $ -0.82\pm 0.10$    &  26.15 &  31.42 &  31.67 &  32.66 &  46.71 &  46.16   &    46.68\\
1133$+$0214 &    1.468 &  Hi & $ -0.71\pm 0.15$    &  25.95 &  31.17 &  31.38 &  32.62 &  46.67 &  46.12 & $< 46.44 $\\
1203$+$1530 &    1.628 &  Hi & $ -0.94\pm 0.07$    &  25.06 &  31.07 &  31.35 &  32.39 &  46.44 &  45.89 & $< 46.82 $\\
1205$+$1436 &    1.643 &  Hi & $ -0.54\pm 0.06$    &  26.93 &  31.19 &  31.35 &  32.48 &  46.53 &  45.98   &    46.99\\
1208$+$1535 &    1.961 &  Hi & $  0.05\pm 0.26$    &  25.87 &  31.33 &  31.32 &  32.64 &  46.69 &  46.14   &    46.85\\
1212$+$1445 &    1.627 &  Hi & $ -0.71\pm 0.14$ & $< 25.49$ &  31.37 &  31.59 &  32.36 &  46.41 &  45.86   &    46.62\\
1216$+$1103 &    1.620 &  Hi & $ -0.51\pm 0.12$    &  25.48 &  31.30 &  31.46 &  32.72 &  46.77 &  46.22 & $< 46.66 $\\
1224$+$1349 &    1.838 &   ? & $ -0.67\pm 0.09$ & $\cdots$  &  31.42 &  31.62 &  32.77 &  46.81 &  46.26   &    46.98\\
1231$+$1320 &    2.380 &  Lo & $ -0.90\pm 0.30$ & $< 25.61$ &  31.65 &  31.92 &  32.61 &  46.66 &  46.10   &    46.87\\
1235$+$0857 &    2.898 &   ? & $ -0.82\pm 0.09$    &  27.39 &  32.04 &  32.29 &  33.37 &  47.42 &  46.87 & $< 46.17 $\\
1235$+$1453 &    2.699 &   ? & $ -0.54\pm 0.12$    &  24.95 &  31.49 &  31.65 &  33.03 &  47.08 &  46.53 & $< 45.83 $\\
1239$+$0955 &    2.013 &  Hi & $ -0.59\pm 0.22$    &  26.20 &  31.51 &  31.69 &  32.83 &  46.88 &  46.33 & $< 45.63 $\\
1240$+$1607 &    2.360 &  Hi & $ -0.72\pm 0.18$    &  25.77 &  31.34 &  31.55 &  32.57 &  46.62 &  46.07 & $< 47.18 $\\
1243$+$0121 &    2.796 &   ? & $ -0.66\pm 0.08$    &  26.10 &  31.71 &  31.91 &  33.31 &  47.36 &  46.81 & $< 46.11 $\\
1314$+$0116 &    2.686 &   ? & $ -0.87\pm 0.05$    &  25.79 &  31.65 &  31.91 &  32.67 &  46.72 &  46.17 & $< 47.10 $\\
1331$-$0108 &    1.881 &  Lo & $ -1.45\pm 0.26$    &  25.57 &  31.61 &  32.05 &  33.03 &  47.08 &  46.53   &    47.43\\
1442$-$0011 &    2.226 &  Hi & $ -0.39\pm 0.22$ & $< 25.78$ &  31.52 &  31.64 &  33.05 &  47.10 &  46.54 & $< 47.34 $\\
1443$+$0141 &    2.451 &   ? & $ -0.52\pm 0.20$    &  25.93 &  31.48 &  31.64 &  32.99 &  47.03 &  46.48 & $< 45.78 $\\
2111$-$4335 &    1.708 &  Hi & $ -0.12\pm 0.02$    &  26.14 &  31.83 &  31.87 &  33.00 &  47.05 &  46.50 & $< 45.80 $\\
2116$-$4439 &    1.480 &  Hi & $ -0.32\pm 0.13$    &  25.00 &  31.34 &  31.44 &  32.63 &  46.68 &  46.13   &    46.37\\
2140$-$4552 &    1.688 &  Hi & $ -0.13\pm 0.04$    &  26.05 &  31.17 &  31.21 &  32.60 &  46.64 &  46.09 & $< 46.68 $\\
2154$-$2005 &    2.035 &  Hi & $ -0.51\pm 0.09$    &  26.10 &  31.45 &  31.61 &  32.97 &  47.02 &  46.47 & $< 45.77 $\\
2201$-$1834 &    1.814 &  Hi & $ -0.59\pm 0.60$ & $< 25.71$ &  31.63 &  31.80 &  32.65 &  46.70 &  46.15   &    46.74\\
2211$-$1915 &    1.952 &  Hi & $ -0.71\pm 0.10$    &  26.82 &  31.53 &  31.74 &  32.62 &  46.67 &  46.12   &    46.56\\
2212$-$1759 &    2.217 &  Hi & $ -0.63\pm 0.07$ & $< 24.68$ &  31.64 &  31.83 &  32.83 &  46.88 &  46.33   &    47.18\\
2350$-$0045A &    1.624 &  Lo & $ -0.63\pm 0.11$ & $< 25.21$ &  31.07 &  31.26 &  32.54 &  46.59 &  46.04 & $< 46.70 $\\
2358$+$0216 &    1.872 &  Lo & $ -1.37\pm 0.49$ & $< 25.65$ &  31.47 &  31.88 &  32.78 &  46.83 &  46.28   &    47.39\\
\enddata
\tablenotetext{a}{Redshifts from \citet{HewFol2003}.}
\tablenotetext{b}{Key: Hi = ultraviolet spectra show high-ionization BALs only; Lo =
low-ionization (\ion{Al}{3} and/or \ion{Mg}{2}) BALs; ? = BAL type unknown because of redshift \citep{LaHu2004,gall+06,clavel+06}.}
\tablenotetext{c}{Spectral index for $l_{\nu}\propto\nu^{\alpha_{\rm
o}}$ from a fit to the rest-frame 1200--5000\AA\ photometry.}
\tablenotetext{d}{Monochromatic 2~keV luminosity converted from the \chandra\ data originally presented in \citet{gall+06}.}
\tablenotetext{e}{The 2500 and 5000\AA\ monochromatic luminosities
are derived from the power-law model fit to the
rest-frame 1200--5000\AA\ photometry.}
\tablenotetext{f}{The 8\micron\ monochromatic luminosity
 is calculated from the \citet{ric+06} composite
luminous quasar SED normalized to the 24\micron\ MIPS data point.}
\tablenotetext{g}{Spectrally integrated infrared (1mm to 2\micron) and
  far-infrared (1mm to 20\micron) quasar luminosities are calculated
  from $\nu$\leight\ as described in \S\ref{sec:seds}.}
\tablenotetext{h}{Spectrally integrated far-infared (1mm to 20\micron) luminosity
  from star formation calculated by normalizing a starburst model to
  the far-infared emission in excess of a composite quasar SED (see
  \S\ref{sec:fir}).}
\end{deluxetable}
\clearpage
\begin{deluxetable}{lrcccccc}
\tabletypesize{\small}
\tablewidth{0pt}
\tablecaption{Infrared, Submillimeter, and Radio Photometry\tablenotemark{a}
\label{tab:phot1}
}
\tablehead{
\colhead{Name} &
\multicolumn{3}{c}{\spitzer\ MIPS} &
\multicolumn{2}{c}{SCUBA\tablenotemark{b}}& 
\multicolumn{2}{c}{Radio}\\
\colhead{(LBQS B)} &
\colhead{24\micron} &
\colhead{70\micron}  &
\colhead{160\micron} &
\colhead{450\micron} &
\colhead{850\micron} &
\colhead{5~GHz\tablenotemark{c}} & 
\colhead{1.4~GHz\tablenotemark{d}} 
}
\startdata
0004$+$0147  & $  7.05\pm  0.13 $ & $< 23.34          $ & $<143.88          $  &           $\cdots$  &           $\cdots$ & $<  0.33          $ & $<  5.00          $\\
0009$+$0219  & $  1.80\pm  0.16 $ & $< 29.44          $ & $<158.84          $  &           $\cdots$  &           $\cdots$  &           $\cdots$ & $<  5.00          $\\
0019$+$0107  & $  8.74\pm  0.18 $ & $< 26.15          $ & $<220.99          $  & $  50.0\pm  16.0 $  & $   8.2\pm   2.3 $ & $<  0.27          $ & $<  0.92          $\\
0021$-$0213  & $  5.57\pm  0.15 $ & $< 33.23          $ & $<203.47          $ & $<  17.4          $  & $   5.3\pm   1.1 $ & $<  0.30          $ & $<  1.00          $\\
0025$-$0151  & $  3.58\pm  0.14 $ & $< 35.68          $ & $<159.29          $ & $<  32.0          $  & $   3.5\pm   1.4 $ & $<  0.36          $ & $<  1.02          $\\
0029$+$0017  & $  6.14\pm  0.17 $  & $ 28.18\pm  7.73 $ & $< 99.77          $ & $<  41.0          $  & $   5.2\pm   2.0 $ & $<  0.27          $ & $<  0.95          $\\
0051$-$0019  & $  8.00\pm  0.15 $ & $< 25.30          $ & $<129.65          $  &           $\cdots$  &           $\cdots$  &           $\cdots$ & $<  0.92          $\\
0054$+$0200  & $  5.12\pm  0.15 $ & $< 20.67          $ & $< 84.68          $  &           $\cdots$  &           $\cdots$  &           $\cdots$ & $<  5.00          $\\
0059$-$2735  & $ 11.27\pm  0.29 $  & $ 34.07\pm  7.24 $ & $\cdots          $  &           $\cdots$  &           $\cdots$ & $<  0.36          $ & $<  5.00          $\\
0106$-$0113  & $  5.83\pm  0.15 $ & $< 29.81          $ & $<497.86          $  &           $\cdots$  &           $\cdots$  &           $\cdots$ & $<  0.96          $\\
0109$-$0128  & $  6.88\pm  0.17 $ & $< 35.20          $ & $<463.49          $  &           $\cdots$  &           $\cdots$  &           $\cdots$ & $<  0.97          $\\
1029$-$0125  & $  4.62\pm  0.18 $  & $ 41.27\pm  7.98 $ & $<171.81          $ & $<  17.2          $ & $<   3.4          $ & $<  0.27          $ & $<  1.14          $\\
1133$+$0214  & $  8.78\pm  0.18 $ & $< 27.83          $ & $<139.10          $  &           $\cdots$  &           $\cdots$  &           $\cdots$ & $<  1.01          $\\
1203$+$1530  & $  4.10\pm  0.16 $ & $< 54.21          $ & $< 88.74          $  &           $\cdots$  &           $\cdots$  &           $\cdots$ & $<  0.95          $\\
1205$+$1436  & $  4.93\pm  0.16 $  & $ 53.69\pm 10.54 $ & $<137.77          $  &           $\cdots$  &           $\cdots$  &           $\cdots$ & $<  0.96          $\\
1208$+$1535  & $  4.78\pm  0.17 $  & $ 43.79\pm  8.08 $ & $< 97.89          $ & $<  34.0          $ & $<   5.0          $ & $<  0.33          $ & $<  0.95          $\\
1212$+$1445  & $  3.80\pm  0.17 $  & $ 22.28\pm  6.65 $ & $<106.44          $  &           $\cdots$  &           $\cdots$  & $  0.46\pm  0.08 $ & $<  0.95          $\\
1216$+$1103  & $  8.78\pm  0.17 $ & $< 40.16          $ & $<100.47          $  &           $\cdots$  &           $\cdots$ & $<  0.36          $ & $<  0.95          $\\
1224$+$1349  & $  7.35\pm  0.15 $  & $ 40.79\pm  8.09 $ & $<117.84          $  &           $\cdots$  &           $\cdots$  &           $\cdots$ & $<  0.97          $\\
1231$+$1320  & $  2.82\pm  0.21 $  & $ 30.13\pm  8.31 $ & $<200.70          $ & $<  16.3          $ & $<   2.8          $ & $<  0.24          $ & $<  1.97          $\\
1235$+$0857  & $ 10.49\pm  0.14 $ & $< 47.89          $ & $< 89.45          $ & $<  13.8          $ & $<   3.4          $  & $  0.51\pm  0.07 $ & $<  1.02          $\\
1235$+$1453  & $  5.63\pm  0.15 $ & $< 26.15          $ & $< 76.00          $ & $<  40.0          $ & $<   4.7          $ & $<  0.39          $ & $<  0.91          $\\
1239$+$0955  & $  6.90\pm  0.14 $ & $< 40.33          $ & $<178.90          $ & $<  10.8          $ & $<   2.0          $  &           $\cdots$ & $<  0.98          $\\
1240$+$1607  & $  2.65\pm  0.15 $ & $< 18.20          $ & $<120.86          $  &           $\cdots$  &           $\cdots$ & $<  0.30          $ & $<  0.98          $\\
1243$+$0121  & $  9.89\pm  0.18 $ & $< 26.11          $ & $<193.57          $ & $<  14.4          $ & $<   3.8          $ & $<  0.30          $ & $<  1.05          $\\
1314$+$0116  & $  2.50\pm  0.14 $ & $< 35.06          $ & $< 87.90          $  &           $\cdots$  &           $\cdots$ & $<  0.27          $ & $<  0.90          $\\
1331$-$0108  & $ 12.89\pm  0.37 $  & $ 76.59\pm  7.97 $ & $\cdots          $  &           $\cdots$  &           $\cdots$  & $  1.54\pm  0.19 $  & $  2.99\pm  0.14 $\\
1442$-$0011  & $  9.04\pm  0.15 $ & $< 50.13          $ & $<174.74          $  &           $\cdots$  &           $\cdots$  & $  0.22\pm  0.06 $ & $<  0.95          $\\
1443$+$0141  & $  6.32\pm  0.13 $ & $< 23.75          $  & $120.00\pm 36.73 $  & $  33.8\pm   8.5 $  & $   5.2\pm   1.2 $ & $<  0.27          $ & $<  1.00          $\\
2111$-$4335  & $ 14.81\pm  0.12 $ & $< 26.51          $  & $117.20\pm 34.82 $  &           $\cdots$  &           $\cdots$  &           $\cdots$   &          $\cdots$\\
2116$-$4439  & $  8.80\pm  0.12 $  & $ 29.58\pm  5.17 $ & $< 90.92          $  &           $\cdots$  &           $\cdots$  &           $\cdots$   &          $\cdots$\\
2140$-$4552  & $  6.01\pm  0.14 $ & $< 20.70          $ & $<198.78          $  &           $\cdots$  &           $\cdots$  &           $\cdots$   &          $\cdots$\\
2154$-$2005  & $  9.34\pm  0.15 $ & $< 26.93          $ & $<102.33          $ & $<  32.0          $ & $<   5.2          $  &           $\cdots$ & $<  5.00          $\\
2201$-$1834  & $  5.78\pm  0.17 $  & $ 30.73\pm  9.95 $ & $<206.85          $ & $<  26.0          $ & $<   6.0          $ & $<  0.33          $ & $<  5.00          $\\
2211$-$1915  & $  4.56\pm  0.14 $ & $< 25.49          $  & $ 69.62\pm 17.08 $ & $<  38.0          $ & $<   4.1          $  &           $\cdots$  & $ 64.00\pm  2.00 $\\
2212$-$1759  & $  5.59\pm  0.14 $  & $ 31.33\pm  8.93 $ & $< 98.11          $  &           $\cdots$  &           $\cdots$ & $<  0.33          $ & $<  5.00          $\\
2350$-$0045A  & $  5.81\pm  0.14 $ & $< 31.38          $ & $<125.70          $  &           $\cdots$  &           $\cdots$ & $<  0.36          $ & $<  1.00          $\\
2358$+$0216  & $  7.31\pm  0.21 $  & $ 79.41\pm 10.49 $ & $<239.25          $  &           $\cdots$  &           $\cdots$  &           $\cdots$ & $<  5.00          $\\
\enddata
\tablenotetext{a}{All flux densities are in units of mJy.}
\tablenotetext{b}{SCUBA data were originally presented in
  \citet{priddey+06}. See \S{\ref{sec:scuba}} for details on detection
criteria and upper limits.}
\tablenotetext{c}{All the 5~GHz data are from \citet{stocke+92}.}
\tablenotetext{d}{Data are compiled from the FIRST \citep{first_ref}
  and NVSS \citep{nvss_ref} catalogs.}
\end{deluxetable}
\clearpage
\begin{landscape}
\begin{deluxetable}{lccccccccccccccc}
\tabletypesize{\footnotesize}
\tablewidth{0pt}
\tablecaption{X-ray, Optical and Near-Infrared Photometry\tablenotemark{a}
\label{tab:phot2}
}
\tablehead{
\colhead{Name} &
\colhead{\chandra\tablenotemark{b}} &
\colhead{LBQS\tablenotemark{c}} &
\multicolumn{5}{c}{SDSS\tablenotemark{d}} &
\multicolumn{3}{c}{2MASS\tablenotemark{e}}\\
\colhead{(LBQS B)} &
\colhead{1~keV} &
\colhead{$B_{\rm J}$} &
\colhead{$u$}  &
\colhead{$g$} &
\colhead{$r$} &
\colhead{$i$} &
\colhead{$z$} & 
\colhead{$J$} & 
\colhead{$H$} & 
\colhead{$K$} 
}
\startdata
0004$+$0147   &    $< 0.246$                    & $ 0.239\pm 0.033 $&           $\cdots$  & $0.234\pm0.032 $  & $0.324\pm0.045 $  & $0.384\pm0.053 $&           $\cdots$  & $0.358\pm0.037 $  & $0.423\pm0.055 $  & $0.384\pm0.063 $\\
0009$+$0219   &    $\cdots$   		        & $ 0.271\pm 0.038 $&           $\cdots$  & $0.273\pm0.038 $  & $0.276\pm0.038 $&           $\cdots$&           $\cdots$  & $0.439\pm0.041 $  & $0.654\pm0.084 $  & $0.373\pm0.078 $\\
0019$+$0107   &    $ 1.150^{+0.437}_{-0.327}$   & $ 0.247\pm 0.034 $&           $\cdots$  & $0.222\pm0.031 $  & $0.294\pm0.041 $  & $0.333\pm0.046 $&           $\cdots$  & $0.446\pm0.038 $  & $0.566\pm0.067 $  & $0.577\pm0.067 $\\
0021$-$0213   &    $ 0.215^{+0.171}_{-0.102}$   & $ 0.144\pm 0.020 $&           $\cdots$  & $0.137\pm0.019 $  & $0.188\pm0.026 $  & $0.198\pm0.027 $&           $\cdots$  & $0.264\pm0.040 $  & $0.355\pm0.056 $ & $<0.471         $\\
0025$-$0151   &    $ 0.466^{+0.280}_{-0.186}$   & $ 0.254\pm 0.035 $&           $\cdots$  & $0.221\pm0.031 $  & $0.267\pm0.037 $  & $0.321\pm0.045 $&           $\cdots$  & $0.525\pm0.042 $  & $0.520\pm0.049 $  & $0.703\pm0.073 $\\
0029$+$0017   &    $ 4.568^{+0.822}_{-0.706}$   & $ 0.149\pm 0.021 $  & $0.066\pm0.002 $  & $0.142\pm0.004 $  & $0.153\pm0.003 $  & $0.158\pm0.005 $  & $0.206\pm0.005 $  & $0.212\pm0.038 $  & $0.274\pm0.052 $  & $0.397\pm0.065 $\\
0051$-$0019   &    $ 2.353^{+0.545}_{-0.452}$   & $ 0.145\pm 0.020 $  & $0.098\pm0.002 $  & $0.152\pm0.002 $  & $0.181\pm0.003 $  & $0.242\pm0.004 $  & $0.279\pm0.006 $  & $0.246\pm0.054 $  & $0.288\pm0.062 $  & $0.315\pm0.069 $\\
0054$+$0200   &    $ 1.173^{+0.387}_{-0.299}$   & $ 0.184\pm 0.026 $&           $\cdots$  & $0.175\pm0.024 $  & $0.213\pm0.030 $  & $0.285\pm0.040 $&           $\cdots$  & $0.187\pm0.037 $  & $0.242\pm0.051 $  & $0.257\pm0.056 $\\
0059$-$2735   &    $< 0.488$		        & $ 0.239\pm 0.033 $&           $\cdots$  & $0.243\pm0.034 $  & $0.371\pm0.051 $  & $0.521\pm0.072 $&           $\cdots$  & $0.704\pm0.043 $  & $1.071\pm0.056 $  & $0.840\pm0.078 $\\
0106$-$0113   &    $\cdots$		        & $ 0.252\pm 0.035 $  & $0.199\pm0.005 $  & $0.235\pm0.004 $  & $0.274\pm0.004 $  & $0.342\pm0.006 $  & $0.349\pm0.007 $  & $0.368\pm0.038 $  & $0.427\pm0.058 $  & $0.308\pm0.069 $\\
0109$-$0128   &    $ 0.654^{+0.299}_{-0.214}$   & $ 0.200\pm 0.028 $  & $0.172\pm0.004 $  & $0.216\pm0.004 $  & $0.233\pm0.003 $  & $0.292\pm0.006 $  & $0.297\pm0.006 $  & $0.355\pm0.039 $  & $0.320\pm0.063 $  & $0.497\pm0.070 $\\
1029$-$0125   &    $ 1.431^{+0.575}_{-0.422}$   & $ 0.181\pm 0.025 $&           $\cdots$  & $0.168\pm0.023 $  & $0.219\pm0.030 $  & $0.293\pm0.041 $&           $\cdots$  & $0.379\pm0.042 $  & $0.434\pm0.051 $  & $0.584\pm0.103 $\\
1133$+$0214   &    $ 1.540^{+0.533}_{-0.409}$   & $ 0.189\pm 0.026 $  & $0.184\pm0.004 $  & $0.249\pm0.003 $  & $0.281\pm0.004 $  & $0.338\pm0.006 $  & $0.327\pm0.008 $  & $0.296\pm0.045 $  & $0.466\pm0.075 $ & $<0.196         $\\
1203$+$1530   &    $ 0.208^{+0.164}_{-0.098}$   & $ 0.141\pm 0.020 $  & $0.090\pm0.002 $  & $0.126\pm0.002 $  & $0.166\pm0.003 $  & $0.216\pm0.004 $  & $0.224\pm0.005 $  & $0.307\pm0.044 $  & $0.382\pm0.064 $  & $0.364\pm0.057 $\\
1205$+$1436   &    $10.400^{+1.310}_{-1.168}$   & $ 0.189\pm 0.026 $  & $0.165\pm0.004 $  & $0.198\pm0.004 $  & $0.212\pm0.004 $  & $0.269\pm0.004 $  & $0.247\pm0.006 $  & $0.334\pm0.038 $  & $0.256\pm0.064 $  & $0.370\pm0.055 $\\
1208$+$1535   &    $ 0.762^{+0.452}_{-0.297}$   & $ 0.287\pm 0.040 $  & $0.137\pm0.004 $  & $0.178\pm0.003 $  & $0.220\pm0.004 $  & $0.253\pm0.005 $  & $0.283\pm0.006 $  & $0.189\pm0.046 $ & $<0.327         $  & $0.280\pm0.061 $\\
1212$+$1445   &    $< 0.463$		        & $ 0.303\pm 0.042 $  & $0.188\pm0.005 $  & $0.278\pm0.005 $  & $0.389\pm0.004 $  & $0.455\pm0.007 $  & $0.445\pm0.008 $  & $0.474\pm0.036 $  & $0.580\pm0.069 $  & $0.390\pm0.062 $\\
1216$+$1103   &    $ 0.551^{+0.333}_{-0.218}$   & $ 0.208\pm 0.029 $  & $0.236\pm0.008 $  & $0.274\pm0.007 $  & $0.308\pm0.005 $  & $0.378\pm0.006 $  & $0.359\pm0.007 $  & $0.395\pm0.059 $  & $0.461\pm0.075 $  & $0.433\pm0.098 $\\
1224$+$1349   &    $\cdots$		        & $ 0.222\pm 0.031 $  & $0.224\pm0.005 $  & $0.227\pm0.005 $  & $0.256\pm0.004 $  & $0.357\pm0.007 $  & $0.378\pm0.008 $  & $0.462\pm0.048 $  & $0.385\pm0.068 $  & $0.394\pm0.058 $\\
1231$+$1320   &    $< 0.308$		        & $ 0.124\pm 0.017 $  & $0.054\pm0.002 $  & $0.187\pm0.004 $  & $0.339\pm0.007 $  & $0.438\pm0.006 $  & $0.503\pm0.011 $  & $0.368\pm0.038 $  & $0.535\pm0.064 $  & $0.791\pm0.066 $\\
1235$+$0857   &    $10.238^{+1.258}_{-1.131}$   & $ 0.230\pm 0.032 $  & $0.039\pm0.002 $  & $0.315\pm0.003 $  & $0.399\pm0.006 $  & $0.520\pm0.008 $  & $0.647\pm0.012 $  & $0.689\pm0.058 $  & $0.911\pm0.071 $  & $0.967\pm0.106 $\\
1235$+$1453   &    $ 0.121^{+0.097}_{-0.059}$   & $ 0.161\pm 0.022 $  & $0.034\pm0.002 $  & $0.122\pm0.002 $  & $0.128\pm0.003 $  & $0.168\pm0.004 $  & $0.210\pm0.007 $  & $0.208\pm0.039 $  & $0.275\pm0.059 $  & $0.362\pm0.059 $\\
1239$+$0955   &    $ 1.730^{+0.531}_{-0.415}$   & $ 0.189\pm 0.026 $  & $0.261\pm0.007 $  & $0.303\pm0.005 $  & $0.318\pm0.004 $  & $0.346\pm0.004 $  & $0.414\pm0.007 $  & $0.400\pm0.058 $  & $0.379\pm0.073 $  & $0.386\pm0.097 $\\
1240$+$1607   &    $ 0.515^{+0.280}_{-0.191}$   & $ 0.124\pm 0.017 $  & $0.046\pm0.002 $  & $0.098\pm0.003 $  & $0.136\pm0.004 $  & $0.150\pm0.003 $  & $0.181\pm0.007 $ & $<0.264         $ & $<0.337         $ & $<0.450         $\\
1243$+$0121   &    $ 0.789^{+0.337}_{-0.247}$   & $ 0.170\pm 0.024 $  & $0.032\pm0.001 $  & $0.203\pm0.003 $  & $0.210\pm0.003 $  & $0.262\pm0.003 $  & $0.330\pm0.006 $  & $0.331\pm0.068 $  & $0.430\pm0.076 $  & $0.445\pm0.081 $\\
1314$+$0116   &    $ 0.321^{+0.252}_{-0.153}$   & $ 0.148\pm 0.020 $  & $0.047\pm0.002 $  & $0.163\pm0.003 $  & $0.196\pm0.003 $  & $0.215\pm0.005 $  & $0.255\pm0.007 $  & $0.377\pm0.048 $  & $0.458\pm0.056 $ & $<0.588         $\\
1331$-$0108   &    $ 0.509^{+0.304}_{-0.201}$   & $ 0.303\pm 0.042 $  & $0.133\pm0.003 $  & $0.215\pm0.004 $  & $0.443\pm0.011 $  & $0.684\pm0.011 $  & $0.732\pm0.016 $  & $0.746\pm0.047 $  & $0.969\pm0.071 $  & $0.860\pm0.092 $\\
1442$-$0011   &    $< 0.513$		        & $ 0.216\pm 0.030 $  & $0.104\pm0.002 $  & $0.198\pm0.004 $  & $0.296\pm0.004 $  & $0.311\pm0.004 $  & $0.352\pm0.007 $  & $0.271\pm0.041 $  & $0.238\pm0.046 $  & $0.493\pm0.083 $\\
1443$+$0141   &    $ 0.657^{+0.356}_{-0.239}$   & $ 0.224\pm 0.031 $  & $0.081\pm0.003 $  & $0.144\pm0.002 $  & $0.175\pm0.003 $  & $0.171\pm0.003 $  & $0.207\pm0.006 $  & $0.215\pm0.052 $  & $0.410\pm0.074 $  & $0.451\pm0.090 $\\
2111$-$4335   &    $ 2.284^{+0.481}_{-0.402}$   & $ 0.907\pm 0.126 $&           $\cdots$  & $0.884\pm0.123 $&           $\cdots$&           $\cdots$&           $\cdots$  & $1.010\pm0.076 $  & $0.991\pm0.102 $  & $1.102\pm0.099 $\\
2116$-$4439   &    $ 0.209^{+0.205}_{-0.114}$   & $ 0.361\pm 0.050 $&           $\cdots$  & $0.334\pm0.046 $  & $0.444\pm0.062 $&           $\cdots$&           $\cdots$  & $0.480\pm0.065 $  & $0.487\pm0.083 $ & $<0.248         $\\
2140$-$4552   &    $ 1.488^{+0.539}_{-0.409}$   & $ 0.204\pm 0.028 $&           $\cdots$  & $0.195\pm0.027 $&           $\cdots$&           $\cdots$&           $\cdots$  & $0.225\pm0.059 $ & $<0.324         $ & $<0.337         $\\
2154$-$2005   &    $ 1.539^{+0.457}_{-0.359}$   & $ 0.241\pm 0.033 $&           $\cdots$  & $0.221\pm0.031 $  & $0.249\pm0.034 $  & $0.265\pm0.037 $&           $\cdots$  & $0.385\pm0.041 $  & $0.445\pm0.053 $  & $0.419\pm0.064 $\\
2201$-$1834   &    $< 0.639$		        & $ 0.320\pm 0.044 $&           $\cdots$  & $0.314\pm0.043 $  & $0.644\pm0.089 $  & $0.987\pm0.137 $&           $\cdots$  & $0.503\pm0.047 $  & $0.390\pm0.082 $  & $0.611\pm0.092 $\\
2211$-$1915   &    $ 6.322^{+0.975}_{-0.853}$   & $ 0.264\pm 0.037 $&           $\cdots$  & $0.249\pm0.034 $  & $0.352\pm0.049 $&           $\cdots$&           $\cdots$  & $0.513\pm0.044 $  & $0.435\pm0.072 $  & $0.658\pm0.072 $\\
2212$-$1759   &    $< 0.041$		        & $ 0.284\pm 0.039 $&           $\cdots$  & $0.251\pm0.035 $  & $0.319\pm0.044 $  & $0.372\pm0.052 $&           $\cdots$  & $0.492\pm0.041 $  & $0.897\pm0.064 $  & $1.459\pm0.074 $\\
2350$-$0045A  &    $< 0.248$		        & $ 0.150\pm 0.021 $  & $0.102\pm0.003 $  & $0.159\pm0.003 $  & $0.173\pm0.003 $  & $0.214\pm0.005 $  & $0.214\pm0.005 $  & $0.238\pm0.041 $  & $0.296\pm0.067 $  & $0.337\pm0.077 $\\
2358$+$0216   &    $< 0.521$                    & $ 0.153\pm 0.021 $&           $\cdots$  & $0.153\pm0.021 $  & $0.370\pm0.051 $  & $0.576\pm0.080 $&           $\cdots$  & $0.532\pm0.044 $  & $0.539\pm0.061 $  & $0.605\pm0.075 $\\
\enddata
\tablenotetext{a}{All flux densities are in units of mJy, except for X-ray
  fluxes which are in units of $10^{-6}$~mJy.}
\tablenotetext{b}{\chandra\ X-ray data were all originally presented by \citet{gall+06} with the exception of 2212$-$1759 observed with \xmm\ and presented by \citet{clavel+06}.}
\tablenotetext{c}{Photometry from the original LBQS survey.  Typical
  survey errors of 0.15 mag have been converted to mJy \citep{lbqs_ref}.}
\tablenotetext{d}{SDSS data are dereddened Data Release 5 PSF
  magnitudes (\altcite{sdss_ref}; http://www.sdss.org/dr5). Data with
  incomplete coverage are synthetic flux densities calculated by
  convolving the SDSS $gri$ filter functions with electronic spectra
  (see \S\ref{sec:sdss} for further details).}
\tablenotetext{e}{Photometry from the 2MASS catalog \citep{2mass_ref} or aperture
  photometry as described in \S\ref{sec:2mass}.}
\end{deluxetable}
\clearpage
\end{landscape}
\begin{deluxetable}{lrr|lrr|lrr}
\tabletypesize{\small}
\tablewidth{0pt}
\tablecaption{Composite SEDs
\label{tab:comps}
}
\tablehead{
\multicolumn{3}{c}{SDSS Comparison Quasars} &
\multicolumn{3}{c}{All BAL~Quasars} &
\multicolumn{3}{c}{HiBAL~Quasars} \\
\colhead{$\nu$\tablenotemark{a}} &
\colhead{$l_{\nu}/l_{\rm 5000\AA}$\tablenotemark{b}} &
\colhead{Number\tablenotemark{c}} &
\colhead{$\nu$\tablenotemark{a}} &
\colhead{$l_{\nu}/l_{\rm 5000\AA}$\tablenotemark{b}} &
\colhead{Number\tablenotemark{c}} &
\colhead{$\nu$\tablenotemark{a}} &
\colhead{$l_{\nu}/l_{\rm 5000\AA}$\tablenotemark{b}} &
\colhead{Number\tablenotemark{c}}
}
\startdata
    13.525 &   1.261 &   8     &     13.509  &  $ 1.210 $  &   8    &      13.512  & $  1.210$  &  8  \\
    13.572 &   1.126 &   7     &     13.532  &  $ 1.225 $  &   7    &      13.548  & $  1.249$  &  7  \\
    13.624 &   1.040 &  10     &     13.565  &  $ 1.136 $  &   9    &      13.591  & $  1.178$  &  8  \\
    13.710 &   1.046 &   8     &     13.607  &  $ 1.114 $  &   8    &      13.619  & $  1.123$  &  2  \\
    13.809 &   0.944 &   7     &     13.664  &  $ 1.173 $  &   6    &      14.558  & $ -0.062$  &  8  \\
    14.012 &   0.856 &   8     &     14.555  &  $ 0.004 $  &   8    &      14.594  & $  0.042$  &  7  \\
    14.060 &   0.636 &   7     &     14.578  &  $-0.024 $  &   7    &      14.640  & $  0.053$  & 10  \\
    14.110 &   0.598 &  10     &     14.611  &  $ 0.011 $  &   9    &      14.674  & $  0.024$  &  7  \\
    14.149 &   0.652 &   9     &     14.646  &  $ 0.069 $  &   7    &      14.705  & $ -0.038$  &  8  \\
    14.189 &   0.502 &   9     &     14.674  &  $ 0.054 $  &  11    &      14.748  & $ -0.034$  &  7  \\
    14.216 &   0.384 &   7     &     14.702  &  $-0.038 $  &  12    &      14.787  & $ -0.027$  &  8  \\
    14.240 &   0.399 &   8     &     14.731  &  $-0.044 $  &   9    &      14.819  & $ -0.038$  &  7  \\
    14.260 &   0.469 &   7     &     14.770  &  $ 0.032 $  &   8    &      14.861  & $ -0.071$  &  7  \\
    14.282 &   0.107 &   7     &     14.799  &  $-0.061 $  &   7    &      14.900  & $ -0.077$  &  7  \\
    14.314 &   0.168 &   9     &     14.817  &  $-0.005 $  &   9    &      14.952  & $ -0.116$  &  7  \\
    14.345 &   0.200 &  11     &     14.838  &  $-0.079 $  &   7    &      15.017  & $ -0.086$  &  9  \\
    14.364 &   0.141 &   9     &     14.861  &  $-0.070 $  &   7    &      15.051  & $ -0.066$  &  7  \\
    14.392 &   0.001 &   7     &     14.899  &  $-0.077 $  &   7    &      15.086  & $ -0.198$  &  7  \\
    14.419 &   0.011 &   8     &     14.943  &  $-0.125 $  &  11    &      15.109  & $ -0.181$  &  7  \\
    14.453 &   0.024 &  11     &     14.981  &  $-0.118 $  &   9    &      15.128  & $ -0.218$  &  7  \\
    14.501 & $-$0.014 &   7     &     15.024  & $ -0.100$  &   8    &      15.168  & $ -0.229$  &  7  \\
    14.546 & $-$0.040 &   8     &     15.045  & $ -0.104$  &   7    &      15.199  & $ -0.241$  &  7  \\
    14.629 & $-$0.159 &   8     &     15.071  & $ -0.174$  &  10    &      15.226  & $ -0.286$  & 11  \\
    14.743 & $-$0.055 &   7     &     15.102  & $ -0.198$  &  15    &      15.242  & $ -0.217$  &  9  \\
    14.911 & $-$0.087 &   7     &     15.123  & $ -0.218$  &   8    &      15.271  & $ -0.214$  &  8  \\
    14.971 & $-$0.080 &   7     &     15.146  & $ -0.247$  &   8    &      15.290  & $ -0.414$  &  7  \\
    15.009 & $-$0.094 &   9     &     15.173  & $ -0.257$  &   8    &      15.315  & $ -0.331$  & 11  \\
    15.037 & $-$0.071 &   7     &     15.199  & $ -0.245$  &   8    &      15.342  & $ -0.354$  &  8  \\
    15.056 & $-$0.086 &   7     &     15.225  & $ -0.276$  &  17    &      15.409  & $ -0.533$  &  7  \\
    15.080 & $-$0.169 &   9     &     15.240  & $ -0.136$  &   8    &      17.797  & $ -5.900$  &  7  \\
    15.113 & $-$0.162 &  11     &     15.256  & $ -0.356$  &  12    &      17.835  & $ -5.469$  &  7  \\
    15.135 & $-$0.183 &  10     &     15.274  & $ -0.367$  &  10    &      17.875  & $ -5.666$  &  7  \\
    15.154 & $-$0.173 &   7     &     15.290  & $ -0.414$  &   7    &      17.902  & $ -5.727$  &  3  \\
    15.187 & $-$0.245 &  10     &     15.315  & $ -0.331$  &  11	 &		&	  &   \\
    15.218 & $-$0.173 &  12     &     15.338  & $ -0.357$  &   7	 &		&	  &   \\
    15.235 & $-$0.139 &   7     &     15.350  & $ -0.430$  &   7	 &		&	  &   \\
    15.268 & $-$0.259 &   9     &     15.374  & $ -0.359$  &   8	 &		&	  &   \\
    15.295 & $-$0.254 &   7     &     15.395  & $ -0.446$  &   7	 &		&	  &   \\
    15.315 & $-$0.226 &   8     &     15.463  & $ -0.777$  &   7	 &		&	  &   \\
    15.337 & $-$0.258 &   7     &     17.795  & $ -6.102$  &   7	 &		&	  &   \\
    15.361 & $-$0.222 &   8     &     17.818  & $ -5.651$  &   7	 &		&	  &   \\
    15.391 & $-$0.276 &   9     &     17.854  & $ -5.465$  &   8	 &		&	  &   \\
    15.421 & $-$0.249 &   8     &     17.892  & $ -5.863$  &   7	 &		&	  &   \\
    15.451 & $-$0.486 &   8     &     17.946  & $ -5.938$  &   6 	 &		&	  &   \\
    15.495 & $-$0.824 &   7     &	     & 		& 	 &		&	  &	      \\
    15.565 & $-$1.880 &   7     &	     & 		& 	 &		&	  &	      \\
    15.669 & $-$1.942 &   2     &             &          &        &              &         &          \\  
\enddata
\tablenotetext{a}{ Logarithm of the rest-frame frequency (Hz) in the
    middle of the bin.}
\tablenotetext{b}{ Logarithm of the ratio of the monochromatic
    luminosity at the given frequency to that at 5000\AA.}
\tablenotetext{c}{Number of photometric data points contributing to
  that frequency bin.}
\end{deluxetable}
\clearpage

\end{document}